\documentclass[a4paper,11pt]{article}
\pdfoutput=1
\usepackage{jheppub}
\usepackage{graphicx}
\usepackage{subcaption}
\usepackage{dsfont}
\usepackage[dvipsnames]{xcolor}
\usepackage{etoolbox}
\usepackage[T1]{fontenc}
\usepackage{amsmath}
\usepackage{amssymb}
\usepackage{bbm}
\usepackage{mathtools}
\usepackage{hyperref}
\usepackage{amsthm}
\usepackage{physics}
\usepackage{tensor}
\makeatletter
\patchcmd{\maketitle}{\@fpheader}{\\}{}{}

\makeatother

\usepackage{babel}

\def\MT#1{{\color{Magenta}{ [#1]}}}

\title{Singularities in 2D and 3D quantum black holes}
\subheader{\begin{flushright}
\texttt{CPHT-RR064.102023}
\end{flushright}}
\author[a,b]{Maciej Kolanowski}
\author[c,d]{and Marija Toma\v{s}evi\'c}

\affiliation[a]{Department of Physics, University of California, Santa Barbara, CA 93106, U.S.A.}
\affiliation[b]{Institute of Theoretical Physics, Faculty of Physics, University of Warsaw, Pasteura 5, 02-093 Warsaw, Poland\looseness=-1}
\affiliation[c]{CPHT, CNRS, \'Ecole polytechnique, Institut Polytechnique de Paris, 91120 Palaiseau, France}
\affiliation[d]{Institute for Theoretical Physics, University of Amsterdam, Science Park 904, 1090 GL Amsterdam,
The Netherlands}

\emailAdd{mkolanowski@ucsb.edu}
\emailAdd{m.tomasevic@uva.nl}

\abstract{We study black holes in two and three dimensions that have spacelike curvature singularities behind horizons. The 2D solutions are obtained by dimensionally reducing certain 3D black holes, known as quantum BTZ solutions. Furthermore, we identify the corresponding dilaton potential and show how it can arise from a higher-dimensional theory. Finally, we show that the rotating BTZ black hole develops a singular inner horizon once quantum effects are properly accounted for, thereby solidifying strong cosmic censorship for all known cases.

}

\begin{document}

\leftline{}

\vskip 2cm

\maketitle

\section{Introduction and summary}
\label{sec:intro}

Spacelike curvature singularities remain one of the least understood concepts in the study of quantum theories of gravity. Curvature singularities in general represent mere signals that our theory---General Relativity---must break down at that point\footnote{The term ``point'' is to be interpreted differently with respect to the type of a curvature singularity in question.}, but some singularities are easier to understand than others. For instance, timelike curvature singularities are expected to play the same role as the singularities found in the Coulomb force equation---they represent some new, underlying degrees of freedom, indescribable within the framework of the theory in question. String theory provides several avenues for resolving timelike singularities; some notable examples include \cite{Dixon:1985jw, Aspinwall:1993yb, Witten:1993yc, Strominger:1995cz}. However, perturbative string theory does not seem to suffice when it comes to spacelike singularities, as first explored in \cite{Horowitz:1989bv}\footnote{Recall, however, that the singularities discussed by Horowitz and Steif did not result from non-singular initial data. This important issue was resolved in a follow-up paper \cite{Horowitz:1990ap} where they indeed show that the energy of string states diverges near a singularity formed from non-singular initial data.}. Spacelike singularities, the moments in time when our theory stops being valid, are also conceptually the most difficult ones to comprehend. Although we have a good grasp on their local structure \cite{Belinsky:1970ew, Belinski:2017fas}\footnote{See also \cite{Witten:2022xxp} on the need for a possible extension beyond the BKL formulation in compactified manifolds.}, their resolution will clearly require some new ideas.
Null singularities, on the other hand, were established by mathematical relativists as generic and stable within the regime of \textit{classical} General Relativity \cite{Ori:1995nj, Luk:2013cqa, Dafermos:2017dbw}. However, there is physical ground to believe these will bend down, usually due to quantum effects, to a spacelike form \cite{Emparan:2021yon, Bousso:2022tdb}. 

Obtaining a description of singularities of any type would represent a significant boost in our understanding of quantum gravity. Our best shot lies within the framework of holography, in which one might hope for a dictionary between the singularity and some more familiar concept on the boundary side. However, our knowledge of the holographic dictionary is still incomplete, especially in bulk dimensions higher than two. The situation is significantly improved for 2D bulk theories. One of the first discussions on this topic can be found in \cite{PhysRevD.44.314}, where a 2D black hole was studied in the context of $c=1$ matrix models\footnote{Even with an exact CFT description, it was not enough to resolve the spacelike singularity.}. Recent work has further illuminated the relationship between 2D bulk theories and matrix model descriptions of the boundary---both perturbatively and non-perturbatively; for a review, see \cite{Mertens:2022irh}. 
This has allowed us to understand quantum gravity in 2D in much greater detail than its higher-dimensional counterparts. The main ingredient is the 1-loop exactness of the bulk theory \cite{Stanford:2017thb}\footnote{This result holds for pure dilaton-gravities. Adding matter is possible \cite{Jafferis:2022wez}, and the matrix description is more involved in that case, but the theory is at best an effective one, due to its UV divergences. One of the salient features of our construction is the amelioration of this effect (albeit not a complete resolution): the 2D theory we will be discussing can be seen as coupled to matter, but since the matter in this case is strongly-coupled with a large number of species, one can apply the holographic principle and perform the reduction from a higher-dimensional theory with no matter; see Sec.~\ref{comments}.}, which has allowed us to make progress in black hole physics, notably regarding black hole microstates and the chaotic nature of horizons \cite{Mertens:2022irh}. 

It is natural to expect that this simplified 2D structure can allow us to understand curvature singularities as well. The holographic map works in the following way: a 2D bulk theory can be written in terms of some dilaton-gravity \cite{Grumiller:2002nm, Grumiller:2021cwg}, with a specific dilaton potential.
Certain dilaton-gravities are known to be dual to matrix integrals, so we can obtain a direct map between solutions in the bulk and the matrix theory describing the boundary \cite{Witten:2020wvy, Maxfield:2020ale, Turiaci:2020fjj, Blommaert:2022lbh}. Even though so far we do not have a map for all possible dilaton potentials, obtaining a dilaton-gravity that contains spacelike curvature singularities presents a first step towards a possible holographic understanding of singularities.


One might wonder why we do not consider higher-dimensional theories whose boundary duals are known exactly, such as the very first example of the AdS/CFT correspondence \cite{Maldacena:1997re}. In fact, there had been some progress made in this direction, most notably in \cite{Fidkowski:2003nf, Festuccia:2008zx, Dodelson:2023vrw} (see also recent work \cite{Horowitz:2023ury}). The main idea is to study boundary correlators that follow complex geodesics that pass close to the singularity, thereby encoding information about it in the relevant two-point function. This analysis, however, is necessarily a semiclassical one, relying on the heat kernel expansion, with a possibility of calculating 1-loop corrections. Even though such correlators might encode some information about the singularity, they cannot tell us anything about its resolution, as that would require non-perturbative data. The main advantage of going to 2D lies precisely in the fact that we can leverage the control over non-perturbative effects in our favor, hopefully shedding light on the resolution of spacelike singularities. 

In this paper, we will initiate the first steps in this direction. We will show how to obtain a 2D theory that allows for spacelike singularities behind the horizon, and consequently, write down the dilaton potential that corresponds to such a theory. Note that while our construction is not the only way to obtain solutions with spacelike singularities in 2D dilaton gravity\footnote{A different construction uses the gas of defects deformation of JT gravity, studied in \cite{Witten:2020wvy, Maxfield:2020ale, Turiaci:2020fjj}. Although these theories are dual to matrix integrals and therefore under good non-perturbative control, it is not clear one can obtain them from higher-dimensional theories which have spacetimes with spacelike singularities behind horizons. We thank Adam Levine and Wayne Weng for discussions on this point.}, the advantage of our approach will lie in the physical motivation for the form of the potential. In other words, we will argue that the form of the potential as obtained in \eqref{dila-pot} is universal since it comes naturally from the backreaction of coupling black holes to matter fields; see Sec.~\ref{sec:qBTZ}. 


Namely, it has been shown that including quantum matter and its backreaction can lead to the ``strengthening'' of the singularity behind the horizon---examples include cases in which the inner (Cauchy) horizon is turned into a proper curvature singularity through the backreaction of quantum fields \cite{Frolov:1991nv, Kay:1996hj, Hollands:2019whz, Emparan:2021yon}. This is a known effect that goes under the name of strong cosmic censorship\footnote{Historically, strong cosmic censorship was postulated due to a blueshift effect classical matter undergoes when approaching the Cauchy horizon \cite{Simpson:1973ua, Poisson:1989zz, Ori:1991zz}. In the following decades, it was found that sometimes this blueshift effect is not enough to turn the whole inner horizon into a singularity, leading to a quasi-regular surface \cite{Hintz:2016gwb, Hintz:2016jak, Dias:2018etb, Luna:2019olw}. However, as we will see in Sec.~\ref{sec:cauchy}, recent work has shown that when quantum fields are included, the horizon must turn into a proper curvature singularity in all cases, regardless of the state and initial conditions. The only outlier so far has been the rotating BTZ black hole, which we resolve in Sec.~\ref{sec:rqBTZ}.}. Given that most dimensional reductions give an AdS$_2$ spacetime with a smooth inner horizon \cite{Kunduri:2007vf, Kunduri:2013gce}, to obtain a singularity in a 2D description, we have to take quantum fields into account\footnote{It should be noted that in AdS$_{d\ge4}$, it is much easier to satisfy strong cosmic censorship conjecture due to the stable trapping of null geodesics \cite{Holzegel:2011uu}; for yet another mechanism, see \cite{Hartnoll:2020rwq}.}. 

This can be easily done within the framework of braneworld holography in which a brane---a lower-dimensional manifold---is embedded into a higher-dimensional AdS spacetime \cite{Randall:1999vf}. The brane is allowed to have asymptotics of any kind; the overall construction does not depend on it \cite{Bueno:2022log}. The braneworld setup is sometimes called double holography since one can employ a lower-dimensional holographic dual to the brane as a boundary (or defect) CFT dual\footnote{In this paper, we will not discuss the role of defect CFTs as holographic duals to branes; for comments on this, see Sec.~7 of \cite{Emparan:2022ijy}.}. 

\subsection*{Obtaining quantum-corrected geometries}
The gravity theory on the brane is obtained by following the same steps as for the holographic renormalization and not taking the cutoff to zero, but leaving it at a finite value; for a short review, see \cite{Bueno:2022log}. This in turn introduces higher-curvature corrections on the brane which disappear as one takes the limit of the brane to the boundary. Additionally, due to its holographic nature, there is a large $N$, strongly-coupled CFT on the brane, coupled to gravity. This CFT also comes with a cutoff since obtaining a brane in the bulk corresponds to integrating out the UV degrees of freedom on the boundary side. Importantly, the CFT leaves its mark on the metric of the brane, incorporating quantum contributions into metric coefficients. These quantum contributions are nothing more than just the standard Schwarzschild (or Kerr) terms when interpreted from the higher-dimensional AdS spacetime, but on the brane, they are endowed with a quantum interpretation. As an example (that we will focus on throughout the paper), the brane (quantum) BTZ $g_{tt}$ coefficient can be written as
\begin{equation}
    H(r) = \frac{r^2}{\ell^2} -1 + \frac{F(M)}{r}, \label{universal}
\end{equation}
where $\ell$ is the AdS cosmological constant, and $F(M)$ some function incorporating quantum effects. Setting $F(M) = 0$ recovers the standard BTZ black hole, but reading the metric in the 4D language, we see that it represents nothing more but the Schwarzschild-AdS black hole metric, with $F(M)$ incorporating its mass. This illustrates the core idea of braneworld holography; the full analysis can be found, for instance, in \cite{Emparan_2020}. Nevertheless, we can already see how this inclusion of quantum effects might alter the interior structure of black holes, and in fact, we will see that it is crucial for the formation of curvature singularities. 

In particular, the second important result in our paper is establishing the strong cosmic censorship conjecture for the rotating BTZ black hole. Similarly to the above construction, one can create a quantum rotating BTZ black hole, which also has its metric coefficients altered. However, these quantum effects turn out not to be enough to lead to a curvature singularity at the inner horizon---in fact, rotating quantum BTZ a priori has the same problem as the standard BTZ black hole since the stress tensor is completely regular at the inner horizon. This led to a conjecture in \cite{Emparan:2020rnp} that additional 1-loop bulk effects (that is, $1/N$ effects on the brane) will lead to a singular inner horizon. Here we show that this is indeed true, and we do it in two ways: first by looking at the 2D geometry and calculating the backreaction of a scalar field on this black hole geometry, and second, by numerically showing in 3D that the inner horizon becomes singular; these results are presented in Sec.~\ref{sec:cauchy}. Note that unlike in the static 2D case, we now need to include an extra scalar to show that the inner horizon is singular; in the static case, the geometry itself is enough.




In Sec.~\ref{sec:qBTZ}, we recall some of the basic properties of quantum BTZ black holes and we show from a lower-dimensional perspective that a singularity persists behind the horizon. 
In Sec.~\ref{sec:cauchy} we show that rotating BTZ black holes with quantum corrections satisfy the strong cosmic censorship conjecture, thereby confirming the claim laid out in \cite{Emparan:2020rnp}. In Sec.~\ref{sec:potential}, we write a dilaton potential that corresponds to a theory of black holes with spacelike singularities, and we show how to obtain such a potential from 3D. We further discuss the 4D reduction and lay out the difficulties one encounters, and their possible resolutions. We finish with a discussion and a brief summary in Sec.~\ref{sec:discussion}, and we lay out some future ideas. In  App.~\ref{sec:energy}, we show that the quantum BTZ solution satisfies the relevant energy conditions, and we write the necessary conditions for the parameters of the rotating case. In App.~\ref{app:appendix}, we provide details regarding dimensional reductions and Weyl rescalings used throughout the bulk of the paper.

\section{Inside quantum black holes}
\label{sec:qBTZ} 

\begin{figure}[h]
    \centering
    \includegraphics[width=.9\textwidth]{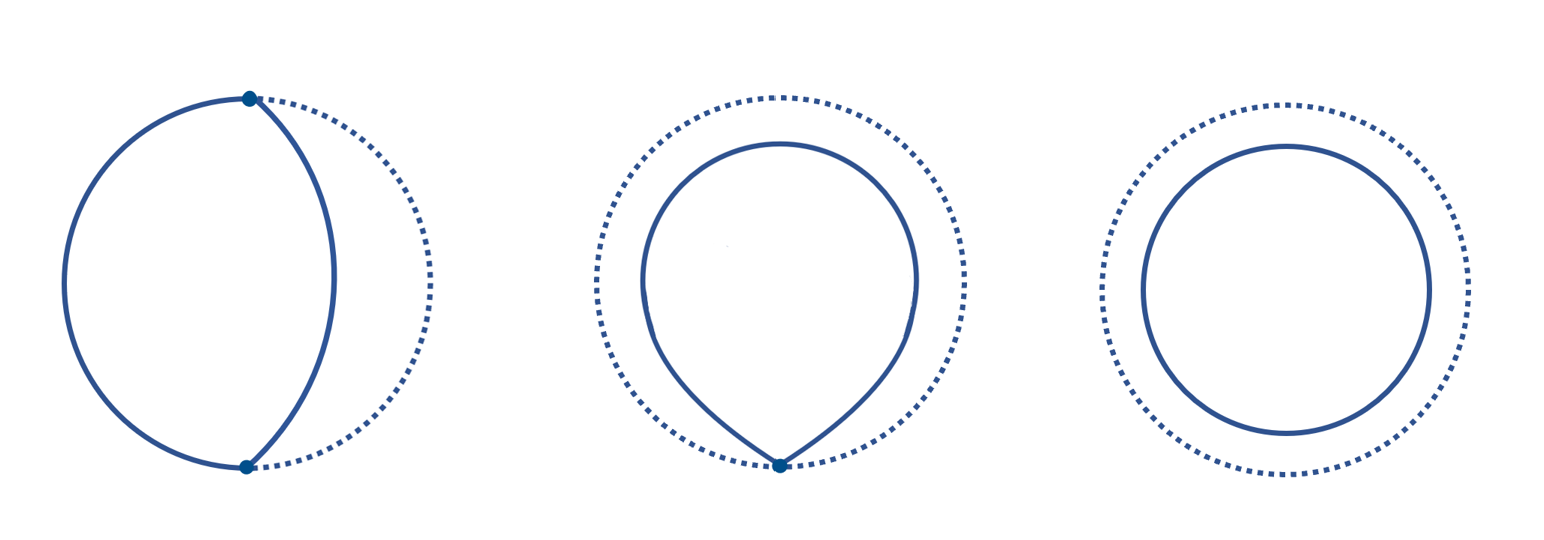}
    \caption{\small{Three types of branes: AdS (left), flat (center), and dS (right). These branes are embedded into a higher-dimensional AdS bulk. The dashed line indicates that the bulk has been integrated out until the full line. The theory on the brane constitutes a lower-dimensional higher-curvature gravity coupled to a CFT with a cutoff. For more details on the construction, see \cite{Emparan_2020}.} All three figures represent branes on a single timeslice, and in this paper, we will always refer to the AdS brane case.}
    \label{fig:branes}
\end{figure}

Let us begin with the dimensional reduction of the quantum-corrected BTZ black hole \cite{Emparan:1999fd, Emparan:2002px, Emparan_2020}. The quantum-corrected BTZ (qBTZ) black hole has been used to verify and analytically obtain several results that depend on quantum-corrected geometries. For instance, one can use this geometry to argue for the non-existence of pathological regions in time machine spacetimes \cite{Emparan:2021xdy, Emparan:2021yon, Tomasevic:2023ojy}, to elucidate the higher-dimensional origin of extended black hole thermodynamics \cite{Frassino:2022zaz}, for the validity of complexity proposals in the presence of quantum corrections \cite{Emparan:2021hyr}, to argue for the existence of quantum black holes in three-dimensional de Sitter and flat spacetimes \cite{Emparan:2022ijy, Panella:2023lsi}, among others. In App.~\ref{sec:energy}, we show that this solution satisfies the average achronal null energy condition---a necessary condition that semiclassical spacetimes must obey. The solution is obtained through a braneworld construction, in which we start with the AdS$_4$ C-metric, and cut the spacetime with an AdS brane. This brane then features a theory of gravity with higher-curvature corrections \cite{Bueno:2022log}, coupled to a CFT with a cutoff; see Fig.~\ref{fig:branes}.


The metric can be written as
\begin{equation}
    ds^2 = -H(r)dt^2 + \frac{dr^2}{H(r)} + r^2d\phi^2, \label{qBTZ}
\end{equation}
where
\begin{equation}
    H(r) = \lambda r^2 + \kappa - \frac{\mu\ell}{r},
\end{equation}
where $\lambda$ is the three-dimensional cosmological constant and different values of $\kappa$ set the global geometry to be spherical, hyperboloidal, or flat. 
Parameters $\mu$ and $\ell$ are related to the parameters of the higher-dimensional bulk (the AdS$_4$ C-metric), but they have an independent braneworld interpretation. Namely, $\mu$ is related to the mass of the higher-dimensional black hole, and, therefore, sets the state of the CFT stress tensor, while $\ell$ controls the tension of the brane, and effectively sets the distance of the brane from the would-be boundary. In braneworld terms (from the 3D theory), however, it controls the strength of the backreaction of the conformal quantum fields. 

The metric as given by \eqref{qBTZ} is not canonically normalized. Indeed, the period of the angle $\phi$ is set by the bulk regularity and in general fails to be $2\pi$. 
The periodicity is given by $\phi\to\phi + 2\pi \Delta$, where $\Delta$ is some constant introduced to avoid a conical singularity that comes from the higher-dimensional metric; more details can be found in \cite{Emparan_2020}. Even though we will be mostly satisfied with the form \eqref{qBTZ}, it will be good to have the canonical metric written down as well. To this end, we can therefore rescale all coordinates appropriately, with
\begin{equation}
    t = \Delta \Bar{t}, \hspace{15pt} r = \frac{\Bar{r}}{\Delta}, \hspace{15pt} \phi = \Bar{\phi} \Delta,
\end{equation}
so that we have
\begin{equation}
    H(\bar{r}) = \lambda \Bar{r}^2 + \kappa \Delta^2 - \frac{\ell \mu \Delta^3}{\Bar{r}}.  \label{qBTZbar}
\end{equation}
Even though the metric obtained seems very specialized, we expect that quantum corrections will lead to the same form of corrections to the metric coefficients. This can be seen through the perturbative calculation of the effect of quantum fields on conical defects in AdS$_3$ \cite{Emparan:2022ijy}, where the calculation can be done explicitly through the method of images, obtaining the same form of the metric as \eqref{qBTZ}.  Curiously, one also obtains the same form of the metric, although with a reduced set of parameters, when coupling a classical conformal massless scalar field to the BTZ black hole \cite{Martinez:1996gn}; we discuss this case more in Sec.~\ref{sec:3D} (see also \cite{Cisterna:2023qhh}). Given that we have several non-trivial examples where the same term $\propto 1/r$ emerges, it seems that the form \eqref{universal} is universal and stems from the backreaction of matter fields.

Note, however, that this statement is true for the static 3D black hole. Once we include rotation, the singularity behind the horizon disappears, and we have a smooth inner horizon, just like for any near-extremal black hole\footnote{Note that there is still the timelike curvature singularity at $r=0$.}. However, as we will see in Sec.~\ref{sec:cauchy}, including extra matter will result in a singular inner horizon once again. One would need to calculate the backreacted metric and see what precise form it would take. Of course, in principle, one can do this calculation for any-D near-extremal black hole: strong cosmic censorship sets the inner horizon to be singular. Moreover, there are reasons to believe that the singularity will always end up being a spacelike one\footnote{We thank Stefan Hollands for discussions on this point.}, as long as one includes enough loops of matter \cite{Bousso:2022tdb}. However, obtaining the exact form of the backreacted metric is a difficult task, and moreover, one would need to include all of the extra matter fields in the action and treat them properly in the dimensional reduction to the 2D dilaton gravity. The static 3D black hole clearly has the advantage of being analytically tractable, while allowing for spacelike singularities to form with some universality, as explained above.

\subsection{The singularity inside (quantum) BTZ} 
The dimensional reduction to the spherical sector is straightforward, as outlined in  App.~\ref{app:dim-red}. Namely, we simply ``peel off'' the circle to obtain
\begin{equation}
     ds^2 = -H(r)dt^2 + \frac{dr^2}{H(r)}. 
\end{equation}
In the case of classical (non-rotating) BTZ, the obtained metric is locally just AdS$_2$. In particular, it does not have any curvature singularity.
The quantum corrections to the BTZ black hole did not affect the spherical symmetry, so one can do the dimensional reduction in the same manner. Note, however, that the metric obtained now is no longer locally AdS$_2$---this is explicitly broken by the quantum corrections $\mu\ell$. Indeed, one can calculate the Kretschmann scalar for this geometry, which is proportional to the square of the scalar curvature $R = -H''(r)$, and so,
\begin{equation}
    R_{\mu\nu\rho\sigma}R^{\mu\nu\rho\sigma} \;\sim\; \frac{\mu^2\ell^2}{r^6}, \label{btzsing}
\end{equation}
which now clearly diverges for $r=0$. Obtaining a curvature singularity for the BTZ black hole is interesting in its own right, but it in fact implies something stronger. The Penrose diagram for the static BTZ black hole is usually depicted as a ``square'' (see below), naively implying that the orbifold singularity is a spacelike one. However, this is false: the orbifold singularity is null, 
and since it is a quasi-regular surface, one can traverse this inner horizon and find closed timelike curves\footnote{One can show this by explicitly mapping the BTZ solution to Misner-AdS$_3$ as in \cite{Emparan:2021yon}.}. In other words, the BTZ singularity is a chronology horizon, so the smoothness of this Cauchy horizon becomes relevant\footnote{Furthermore, one could send signals to the beyond-inner-horizon CFTs, connecting the two boundary theories through a bulk, even though the boundary CFTs are not coupled in any way---this would constitute a violation of the no-transmission principle \cite{Engelhardt:2015gla}.}. 


In fact, we can make a broader conjecture. The singularity found in \eqref{btzsing} came from the additional factor of $1/r$ in \eqref{qBTZbar}. And even though the quantum BTZ black hole is a specific solution, we have already mentioned that this specific form emerges whenever one calculates the backreacted metric after coupling to matter fields (quantum or classically conformal). In other words, this form seems to be a \textit{universal} response to the matter fields propagating in the 3D static black hole background, and this universality is not affected by the dimensional reduction to the AdS$_2$ sector. Given that the emergence of the AdS$_2$ region is inevitable in the extremal limit of many black holes \cite{Kunduri:2007vf, Kunduri:2013gce}, we can make a simple, yet powerful conjecture: since we expect quantum corrections to play a role for all black holes that have an emergent AdS$_2$ sector, this divergence implies that strong cosmic censorship will hold in all known such cases\footnote{Of course, there are cases, such as the Kerr black hole, where one obtains a warped AdS$_2$ geometry, which then implements additional effects. Nevertheless, we already know that Kerr-like solutions obey strong cosmic censorship, so we can see that the warping factor does not invalidate the censorship conjecture and the formation of singularities.}.

\subsection{The Penrose diagram is no longer a square} The existence of a proper curvature singularity also implies that the causal structure of the quantum BTZ black hole is no longer as simple as before. In \cite{Fidkowski:2003nf}, the authors looked at radial null geodesics inside black holes to determine the time it takes to get to the singularity. For BTZ, it was found that the real part of that time goes to zero, indicating that geodesics thrown from the left and the right asymptotic regions meet right in the middle of the singularity. In other words, the Penrose diagram is a ``square''. The intuition is that ``time becomes space'' inside the black hole, so the zero distance in time would indicate that the null geodesics meet at the same point. The calculation is a little bit subtle, due to the horizon pole one needs to take care of, so we will reproduce their result for the BTZ black hole here. 

\paragraph{The BTZ case.} The metric reads easily, where $r_h$ is the location of the event horizon,
\begin{equation}
    ds^2 = -f(r) dt^2 + \frac{dr^2}{f(r)} + r^2 d\varphi^2, \hspace{15pt} f(r) = r^2 - r_h^2,
\end{equation}
and to obtain null geodesics, we simply have to solve
\begin{equation}
    t(R) = t_0 \pm \int_R^\infty \frac{dr}{f(r)},
\end{equation}
where $R$ is some point inside the black hole, $R < r_h$. The $\pm$ refers to ingoing and outgoing light rays, and we will choose the minus sign for the ingoing ones. Now, we need to take care of the pole that exists at $r = r_h$, and we can do this through a slightly complex calculation: namely, our integral over the real line will now make a little jump into the complex plane so that we have
\begin{equation}
    \int_R^\infty = \int_R^{r_h-\epsilon} + \int_{r_h-\epsilon}^{r_h+\epsilon} + \int_{r_h+\epsilon}^\infty,
\end{equation}
where for the second integral we choose $r = r_h + \epsilon e^{i\theta}$ to obtain 
\begin{equation}
    \int_\pi^0 \frac{i\epsilon e^{i\theta} d\theta}{(r_h + \epsilon e^{i\theta})^2 - r_h^2} = -\frac{i\pi}{2r_h},
\end{equation}
and this imaginary contribution exactly tells us that a pole has been accounted for. The other two integrals are simple to solve, and so we obtain as the final result
\begin{equation}
    t(R) = t_0 + \frac{1}{r_h} \text{arctanh}\left(\frac{R}{r_h}\right) - \frac{i\pi}{2r_h}.
\end{equation}
Note that for $R\to 0$, $t(0) = t_0 - \frac{i\pi}{2r_h}$, so we see that the real value of the time $t(0)$ disappears. This is indicative of the fact that null geodesics sent from the left and right asymptotic regions meet at the conical singularity. We will see below that this is no longer the case. 

\paragraph{The quantum BTZ case.} The idea will be the same, but now we will have a more complicated integral to solve. We will write $H(r)$ from \eqref{qBTZ} as
\begin{equation}
    H(r) = \frac{(r-r_1)(r-r_2)(r-r_3)}{r} = r^2 - r_1^2 + \frac{\mu\ell}{\lambda}\frac{r-r_1}{rr_1},
\end{equation}
where $r_i$ are the solutions to $H(r)=0$, and $r_1$ is taken to be the event horizon radius. In this form, we can now easily solve the integral 
\begin{equation}
    \int \frac{dr}{H(r)} = \alpha \gamma_1 \arctan{\left(\alpha (2r+r_1)\right)} + \gamma_2\left[2\log{(r-r_1)} - \log{(m + rr_1(r+r_1))}\right],
\end{equation}
where
\begin{equation}
    \alpha = \sqrt{\frac{r_1}{4m - r_1^3}}, \hspace{10pt} \gamma_1 = \frac{2m+r_1^3}{m+2r_1^3}, \hspace{10pt} \gamma_2 = \frac{r_1^2}{2(m+2r_1^3)}, \hspace{10pt} m \equiv \frac{\mu\ell}{\lambda}.
\end{equation}
We will use this form to evaluate the integrals around the pole, and for the pole, we need to evaluate its residue,
\begin{equation}
    \int_{r_1-\epsilon}^{r_1+\epsilon} \frac{dr}{H(r)} = -\frac{i\pi r_1^2}{m+2r_1^3} = - 2i\pi \gamma_2.
\end{equation}
The full integral is then equal to\footnote{In the case that $4m < r_1^3$, the result changes by a little bit: there is a factor of $i$ multiplying $\pi/2$, and the $+\arctan$ becomes $-\text{arctanh}$ with the same argument. On top of this, one assures that $\alpha(2R + r_1) > 1$ so that the imaginary parts cancel and one is left with the same imaginary signature as in the case $4m > r_1^3$. In this case then, the limit $m\to0$ gives the BTZ result. }
\begin{equation}
\begin{split}
    \int_R^\infty \frac{dr}{H(r)} &= \alpha \gamma_1\left(\frac{\pi}{2} - \arctan{\left(\alpha(2R + r_1)\right)}\right) - 2i\pi\gamma_2\\
    &- \gamma_2\left[\log{r_1} + 2\log{(r_1 - R)} - \log{(m + Rr_1(R+r_1))}\right],
\end{split}
\end{equation}
which for $R\to0$ gives
\begin{equation}
    t(0) = t_0 + \alpha\gamma_1\left(\frac{\pi}{2} - \arctan{\left(\alpha r_1\right)}\right) + \gamma_2 (\log{m} - 3\log{r_1}) - 2i\pi \gamma_2.
\end{equation}
We see that we have a non-zero real part and hence, the time to the singularity will not be fully symmetric from the left and the right wedge as for the BTZ case.

\section{Inner Cauchy horizons}
\label{sec:cauchy}

In this section, we will show that the quantum stress tensor generically diverges at the inner horizon of the two-dimensional geometry. We start with a review of the Reissner-Nordstrom de Sitter geometry and its two-dimensional throat description. We then show what are the main apparent difficulties with the analysis when applied to the dimensionally reduced metric, and we show that once higher corrections are taken into account, the inner horizon will become singular in a similar manner---solidifying the strong cosmic censorship conjecture for all known cases.

Having done that, we move to the full three-dimensional quantum rotating BTZ. We put on that background a probe quantum field. For concreteness, we restricted ourselves to the scalar fields. Then, we obtain the behavior of the expectation value of the energy-momentum tensor close to the Cauchy horizon, using the formalism put forward in \cite{Hollands:2019whz}. The calculation is quite non-trivial due to the issues connected with the global (vs. local) structure of our spacetimes and associated prescription for the prescription of the field's state. When the dust settles, one finds that for generic parameters (of the black hole and the field) we have
\begin{equation}
    \langle \hat{T}_{VV} \rangle \sim V^{-2},
\end{equation}
thus proving the strong cosmic censorship in that background.
\subsection{2D analysis}
\label{sec:cauchy2}

\subsubsection*{Review of the RN-dS analysis}
\label{sec:review} 
Let us first review the two-dimensional analysis of Hollands, Wald, and Zahn from \cite{Hollands:2019whz}; see also \cite{Birrell78, Shrivastava:2020xmw, Hollands:2020qpe, Bhattacharjee:2020nul}. We will start with a two-dimensional metric,
\begin{equation}
    ds^2 = g_{\mu\nu} dx^\mu dx^\nu = -f(r) dt^2 + \frac{dr^2}{f(r)}, \label{metrika}
\end{equation}
where $f(r)$ is the blackening factor; it does not really matter what the exact function is for this analysis. We will only note that it has zeros for any Killing horizons that may be present. In \cite{Hollands:2019whz}, they studied the Reissner-Nordstrom solution in de Sitter space, so the zeroes are given by $r_+, r_-$ and $r_c$, where $c$ stands for the cosmological horizon, and the $\pm$ for the outer and inner horizons. We can also define their respective surface gravities,
\begin{equation}
    \kappa_x = \frac{1}{2} |f'(r_x)|,
\end{equation}
where $x$ stands for any aforementioned horizon.

On top of this spacetime, there is a massless, real scalar field that obeys
\begin{equation}
    \Box \Phi = g^{\mu\nu} \nabla_\mu \nabla_\nu \Phi = 0.
\end{equation}
Recall that in 2d, massless scalar field is also conformal. We will first show that the classical stress tensor does not diverge strongly enough for a certain range of parameters, and then we will show how quantum effects amplify the divergence at the inner horizon. For this analysis, we will mostly work with the Eddington-Finkelstein (EF) coordinates and the Kruskal extension. We define the tortoise coordinate in the following way:
\begin{equation}
    dr_* = \frac{dr}{f(r)},
\end{equation}
and we use it to define the EF coordinates,
\begin{equation}
    u = t - r_*, \hspace{15pt} v = t + r_*,
\end{equation}
which gives us the metric in null coordinates,
\begin{equation}
    ds^2 = - f(r) du dv.
\end{equation}
We can also define the appropriate Kruskal extension, of which we will only need the following:
\begin{equation}
    V_- := - e^{-\kappa_- v}, \hspace{15pt} V_c := -e^{-\kappa_c v}. \label{vs}
\end{equation}
We only need the $V$-extension since we will be looking at trajectories that are transverse to the inner horizon in the left wedge. 
\paragraph{Classical stress tensor.} The classical stress tensor analysis is simple: we only need the conservation equation, $\nabla^\mu T_{\mu\nu} = 0$ and the fact that the classical stress tensor is traceless, $g^{\mu\nu} T_{\mu\nu} = 0$. From these two equations, we can derive the $vv$-component which is given by 
\begin{equation}
    g^{uv}\left(\nabla_v T_{uv} + \nabla_u T_{vv}\right) = 0,
\end{equation}
and since $T_{uv} = 0$, we have
\begin{equation}
    \partial_u T_{vv} = 0,
\end{equation}
that is, our stress tensor is constant along $u$-trajectories. In other words, we have
\begin{equation}
    T_{vv} (U, v) = T_{vv} (U_0, v).
\end{equation}
We now want to take the limit $v\xrightarrow{}\infty$, since this is where the inner horizon will be (as well as the cosmological horizon). In this limit, $T_{vv} (U,v)$ will approach the inner horizon, and $T_{vv}(U_0, v)$ will approach the dS horizon (see Fig.~5 from \cite{Hollands:2019whz}). To obtain the stress tensors in this limit, we will switch to coordinates which are regular across these horizons, that is $V_-$ and $V_c$. Since
\begin{equation}
    T_{\mu\nu} = T_{\alpha\beta} \frac{\partial x^\alpha}{\partial x^\mu} \frac{\partial x^\beta}{\partial x^\nu},
\end{equation}
and through \eqref{vs}
\begin{equation}
    dV_- = -\kappa_- V_- dv, \hspace{15pt} dV_c = -\kappa_c V_c dv = -\kappa_c (-V_-)^{\frac{\kappa_c}{\kappa_-}} dv,
\end{equation}
we have
\begin{equation}
    T_{vv}(U,v) = T_{V_- V_-} \left(\frac{\partial V_-}{\partial v}\right)^2 = T_{V_- V_-} \kappa_-^2 V_-^2, \label{vminus}
\end{equation}
\begin{equation}
    T_{vv}(U_0, v) = T_{V_c V_c} \left(\frac{\partial V_c}{\partial v}\right)^2 = T_{V_c V_c} \kappa_c^2 V_c^2,
\end{equation}
so when we equate them, we obtain
\begin{equation}
    T_{V_- V_-} = \frac{\kappa_c^2}{\kappa_-^2} (- V_-)^{\frac{2\kappa_c}{\kappa_-} - 2} T_{V_c V_c}.
\end{equation}
In order to have a proper impassable divergence at the inner horizon, the divergence in the stress tensor must be at least of order 2. Here we see that the strength of the divergence might be smaller, depending on the ratio of $\kappa_c$ and $\kappa_-$; this is why we will resort to the quantum analysis to see if the strength of the divergence improves.

\paragraph{Quantum stress tensor.} The only difference compared to the previous analysis is in the trace of the stress tensor. In 2D, the trace is famously related to Hawking radiation, and it, in fact, encompasses all of the information regarding Hawking radiation. In other words, the trace is all we need in order to specify our quantum effects.

The trace is given by
\begin{equation}
    g^{\mu\nu} T_{\mu\nu} = \alpha R,
\end{equation}
where $R$ is the Ricci scalar of the metric $g_{\mu\nu}$, and $\alpha$ is a parameter that defines what type of a quantum field propagates on this spacetime; for the massless scalar that we are studying here, $\alpha = \frac{1}{24 \pi}$. The Ricci scalar, in our case, is given by
\begin{equation}
    R = -f'', \hspace{15pt} ' := \frac{d}{dr},
\end{equation}
so from the trace equation, we obtain
\begin{equation}
    T_{uv} = \frac{\alpha}{4} f f'',
\end{equation}
where one factor of 2 comes from the symmetricity of the $uv$-component $2 g^{uv}T_{uv}$, and another from $g_{uv} = -\frac{1}{2} f$. Then, the $\nu = v$ component of the conservation equation gives us
\begin{equation}
    \partial_u T_{vv}  = - f \partial_v \left( f^{-1} T_{uv}\right) = -\frac{\alpha}{8} f^2 f'''.
\end{equation}
To integrate this equation, we will switch to the $r$-coordinate, and so since 
\begin{equation}
    dr_* = \frac{\partial r_*}{\partial u} du + \frac{\partial r_*}{\partial v} dv = \frac{1}{2} \left(-du + dv\right),
\end{equation}
we have
\begin{equation}
    \frac{\partial}{\partial u} = \frac{\partial r_*}{\partial u} \frac{\partial r}{\partial r_*} \frac{\partial}{\partial r} = -\frac{1}{2} f \frac{\partial}{\partial r},
\end{equation}
which gives 
\begin{equation}
    T_{vv}(U,v) - T_{vv}(U_0, v) = \frac{\alpha}{4}\int dr f f''', 
\end{equation}
which can be written as 
\begin{equation}
    T_{vv}(U,v) - T_{vv}(U_0, v) = \frac{\alpha}{4} \left(\int d(ff'') -  \int f' f'' \right) = \frac{\alpha}{4} \left(\int d(ff'') -  \int d(f'^2) + \int f'f'' \right),
\end{equation}
and so finally,
\begin{equation}
    T_{vv}(U,v) = \frac{\alpha}{8}\left(2 ff'' - f'^2\right)|_{r(U_0, v)}^{r(U,v)} + T_{vv}(U_0, v). \label{stress}
\end{equation}
Note that in the limit $v\to\infty$, $r(U,v) = r_-$ and $r(U_0, v) = r_c$, and the blackening factor vanishes for both of these points. However, the first derivative will give us the corresponding surface gravities, and so,
\begin{equation}
    \lim_{v\to\infty} T_{vv}(U,v) =     \lim_{v\to\infty} T_{vv}(U_0, v) + \frac{\alpha}{2}(\kappa^2_c - \kappa^2_-).
\end{equation}
The first term is the same as in the classical case, but the second term is more interesting and will give us the needed behavior at the inner horizon. Transforming to the regular $V$ coordinates \eqref{vminus}, the second term gives us
\begin{equation}
    T_{V_-V_-} = \frac{\alpha}{2}\frac{\kappa^2_c - \kappa^2_-}{\kappa^2_-} \frac{1}{V_-^2}. \label{stressten}
\end{equation}
We see the stress tensor will be divergent unless $\kappa_c = \kappa_-$, and this divergence is present regardless of the possible divergence in the classical term; in other words, it is a universal behavior that emerges in the near horizon limit for quantum fields.

\paragraph{Inner horizons in the AdS$_2$ throat.}

We now want to study two-dimensional geometries in AdS spacetimes, not dS. The main difference with respect to the previous analysis comes in the form of boundary conditions: we no longer have $r_c$ nor a notion of a cosmological horizon, and so we must resort to a different boundary condition. A natural candidate is the outer horizon $r_+$. In the limit, $v\to\infty$, $r(U,v) = r_+$, and the rest follows in the same manner as before. Therefore, we can simply replace $\kappa_c$ in all of our previous results with $\kappa_+$. Unfortunately, in the case of AdS$_2$ geometry, the inner horizon and the outer horizon have the same surface gravity, given that $r_+ = r_0$ and $r_- = -r_0$ \cite{Moitra:2020ojo}. This would then make the entire geometry smooth, even at the classical level! However, one should bear in mind that AdS$_2$ is not the full description of the near-horizon region: one has effects coming from the transverse spheres, as well as from the fields outside of the black hole. All of these effects will make the metric of the nearly-AdS$_2$ region different than the pure AdS$_2$ metric. Also, the horizons will no longer coincide up to a minus sign: this was true for AdS$_2$ since the $g_{tt}$ factor goes as $r^2 - r_0^2$; additional corrections would then lead to different solutions.

\subsubsection*{The rotating quantum BTZ solution}
\label{sec:rqBTZ}

The only exception to the conclusion reached in Sec.~\ref{sec:qBTZ} is the case of a rotating BTZ black hole, which has shown to be an unusually interesting example. Namely, this black hole has a high degree of symmetry, which leads to some non-trivial cancellations in the calculations of \cite{Hollands:2019whz, Papadodimas:2019msp}. These authors considered the entanglement structure across the outer and inner horizons and found miraculous cancellations that led to a smooth inner horizon for the rotating BTZ black hole\footnote{These miraculous cancellations occur both at the classical and quantum levels. However, it is only for certain parameter ranges that the rotating BTZ black hole violates the \textit{classical} strong cosmic censorship \cite{Dias:2019ery}.}. 

It is natural, then, to consider the quantum-corrected version of this black hole to see if the situation improves. In \cite{Emparan:2020rnp}, the authors mapped the quantum-corrected solution to the Kerr-AdS black hole, through the use of double holography, and showed that the two black holes share the inner horizon. Since the Kerr-Ads black hole obeys the strong cosmic censorship, this implies that the rotating quantum BTZ black hole will likewise obey it. Here, we will show this explicitly.

The metric of the rotating quantum BTZ black hole can be locally written as 
\begin{equation}
    ds^2 = -H(r)dt^2 + \frac{dr^2}{H(r)} + r^2\left(d\phi - \frac{a}{r^2}dt\right)^2, \label{rqBTZm}
\end{equation}
where
\begin{equation}
    H(r) = \lambda r^2 + \kappa - \frac{\mu \ell}{r} + \frac{a^2}{r^2}, \hspace{10pt} \lambda = \ell_3^{-2}.
\end{equation}
We can write the blackening factor as 
\begin{equation}
    H(r) = \frac{(r - r_1)(r - r_2)(r - r_3)(r - r_4)}{\ell_3^2 \;r^2} = \frac{\lambda r^4 + \kappa r^2 - \mu \ell r + a^2}{r^2},
\end{equation}
where $r_i$ are the roots of $H(r) = 0$. Two of them, say $r_3$ and $r_4$ are unphysical, so we will rewrite them in terms of the outer $r_1$ and inner $r_2$ horizons. For this, we can use Vieta's formula for a quartic polynomial,
\begin{equation}
    r_1 + r_2 + r_3 + r_4 = 0,
\end{equation}
\begin{equation}
    r_1\cdot r_2\cdot r_3 \cdot r_4 = \frac{a^2}{\lambda},
\end{equation}
\begin{equation}
    r_1 r_2 + (r_1 + r_2)(r_3 + r_4) + r_3 r_4 = \frac{\kappa}{\lambda},
\end{equation}
\begin{equation}
    r_1 r_2 (r_3 + r_4) + r_3 r_4 (r_1 + r_2) = \frac{\mu \ell}{\lambda},
\end{equation}
in order to obtain
\begin{equation}
    H(r) = \frac{(r^2 - r_+^2)(r^2 - r_-^2)}{\ell_3^2 \; r^2} + \frac{\mu \ell}{r^2} \frac{(r - r_+)(r - r_-)}{(r_+ + r_-)},
\end{equation}
where $r_{1,2} = r_\pm$, and we can recognize the first term as the BTZ blackening function\footnote{Setting $r_- = 0$ gives the static case.}. The surface gravity $\kappa_\pm = \frac{1}{2} \abs{H'(r_\pm)}$ is then given by
\begin{equation}
    \kappa_\pm = \frac{1}{2} \frac{(r_+ - r_-)(2 r_\pm (r_+ + r_-)^2 +\lambda \mu \ell)}{\lambda r_\pm^2 (r_+ + r_-)}.
\end{equation}
We see that in the extremal limit $r_+ \to r_-$ the surface gravity, and so the temperature, goes to zero, as for the standard BTZ black hole. 
\noindent To see the emergent AdS$_2$ region, let us perform the same rescaling as in \cite{Ghosh_2020}, where 
\begin{equation}
    r = \frac{1}{2}(r_+ + r_-) + \frac{1}{2} (r_+ - r_-) \rho,
\end{equation}
so we will expand the final result around the extremal limit $r_+ \to r_-$. In that case,
\begin{equation}
    H_{ext}(r) = -\left(1 + \frac{\mu \ell}{8 r_-^3 \lambda}\right) (\rho^2 - 1)(r_+ - r_-)^2.
\end{equation}
We see that the scaling is the same as for rotating BTZ, which sets these quantum corrections on a second-order level---the strong cosmic censorship seems to fail for this geometry.

We can go higher in the order of the extremality parameter, $\xi = r_+ - r_-$. First, we can see that in the standard BTZ case, we obtain\footnote{We thank the authors of \cite{Bhattacharjee:2020nul} for pointing out an error in the previous version of this paper.}
\begin{equation}
     \kappa_\pm = \frac{(r_+^2 - r_-^2)}{ r_\pm\ell_3^2} = \frac{\xi (2r_- \pm \xi)}{r_- \ell^2_3}, \label{btzT}
\end{equation}
and this, of course, indicates that the temperature is different for both horizons and the strong cosmic censorship is satisfied. This is an (apparent) contradiction to the three-dimensional theory. One needs to keep in mind that a given field living in higher dimensions corresponds to the infinite tower of Kaluza-Klein modes of different masses living in 2D. Generically, each of them close to the horizon will have the same asymptotics
\begin{equation}
    \langle \hat{T}_{VV} \rangle \sim V^{-2}.
\end{equation}
However, the proportionality constant will be mass-dependent (but otherwise universal, in particular state independent). The proper way to interpret 3D results from the 2D lense is to acknowledge that contributions from all the Kaluza-Klein modes cancel out exactly. Let us emphasize that this is a highly non-generic situation and we do not expect to encounter that in dimensions higher than three.

Even though we learned that 2D results should be taken with a grain of salt, for the sake of completeness, let us discuss the temperatures of quantum rotating BTZ. Let us take a closer look at the near-extremal solution. In this case, we will send $r_+\to r_- + \xi$, where $\xi \ll 1$, and we will expand in $\xi$ keeping higher-order corrections.  
\begin{equation}
    H'(r_+) = \frac{\xi  \left(-5 l \mu  \xi +2 l \mu  r_-+16 \lambda  r_-^4-8 \lambda  \xi  r_-^3\right)}{4 \lambda  r_-^4},
\end{equation}
while
\begin{equation}
H'(r_-) = -\frac{\xi  \left(-l \mu  \xi +2 l \mu  r_-+16 \lambda  r_-^4+8 \lambda  \xi  r_-^3\right)}{4 \lambda  r_-^4}.
\end{equation}

Let us mention the recent results of \cite{Ghosh_2020, Iliesiu:2020qvm} that indicate that the extremal limit must be taken with care and that quantum effects dominate for such cases, leaving the physical picture of the horizon and its smoothness fuzzy. Thus, we expect further quantum correction to the geometry in this regime. 

Finally, let us write down the geometry in the throat while keeping higher-order terms. The standard rotating BTZ case gives
\begin{equation}
    H_{\text{ext}}^{\text{BTZ}} = \frac{\lambda(r_+ - r_-)^2 (\rho^2 - 1)}{2r_-} \left(r_-(2+\rho) - r_+\rho\right),
\end{equation}
while quantum rotating BTZ gives
\begin{equation}
     H_{\text{ext}}^{\text{qBTZ}} = \frac{\lambda(r_+ - r_-)^2 (\rho^2 - 1)}{16r_-^4} \left(16 r_-^4 + \frac{5\mu\ell r_-}{\lambda} - \frac{3\mu\ell r_+}{\lambda} + 2(r_- - r_+)(4r_-^3 + \frac{\mu\ell}{\lambda})\rho\right).
\end{equation}
We see that both deviate from the AdS$_2$ geometry once higher orders in $\xi = r_+ - r_-$ are taken into account, and this is expected: the AdS$_2$ geometry is supposed to emerge only in the (near-)extremal limit. This is also consistent with the fact that the temperatures deviate as well once we go beyond the leading order limit. 


Given that this analysis was done in 2D, where we neglected the modes coming from the reduced circle, one could argue that perhaps there would be some higher-dimensional ``conspiracy'' that would lead to a smooth inner horizon again, similar to the standard BTZ case. Additionally, the form of the metric we used was the local one (with ``unbarred coordinates''), which does not have well-defined identifications at infinity. Given that the main result depends on the ratio between surface gravities (that do depend on the asymptotics), one might object that our temperatures are not properly obtained. We close these loopholes in the next subsection, where we show explicitly from the 3D perspective that the inner horizon cannot be smooth.

\subsection{3D analysis}
\label{sec:perturbing}
\label{sec:qperts}

In this section, we will show explicitly in 3D that quantum effects lead to a singular inner horizon of the rotating BTZ black hole. Before that, let us mention an observation made in \cite{Dias:2019ery}. It was observed that for a {\it generic} deformation of the background geometry, the classical cancellations are lifted and Christodoulou’s version of strong cosmic censorship \cite{Christodoulou:2008nj} holds for a massless scalar. Nevertheless, it will be violated close to the extremality by massive fields. Thus, even in this case one needs to invoke quantum effects. We will use the method outlined in \cite{Hollands:2019whz}, which we review below.


\begin{figure}[h!]
    \centering    \includegraphics[width=.4\textwidth]{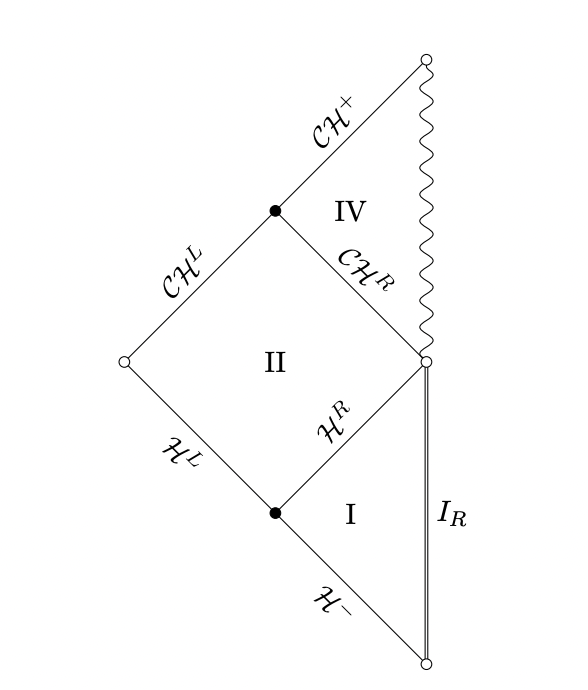}
    \caption{\small{The Penrose diagram of the rotating BTZ black hole. The relevant Cauchy horizon, on which the matter stress tensor is evaluated, is denoted by $\mathcal{CH}^R$}. The rotating quantum BTZ solution has the same causal structure.}   \label{fig:Penrose_diagram_BTZ}
\end{figure}

Consider a Hadamard state $\Psi$ of the scalar field $\Phi$ that is well-defined in regions II and I and another comparison state $C$ that is well-defined in II and IV (see Fig.~\ref{fig:Penrose_diagram_BTZ}). For simplicity, we will assume that the field is minimally-coupled and its mass is $\mu_{\Phi}$. The qualitative answer should not depend on these details of the theory. The energy-momentum tensor of the associated state can be formally obtained from the two-point function:

\begin{equation}
    \langle T_{\mu \nu}(x) \rangle_\Psi :=\lim_{x' \to x} \left(
    \nabla_{\mu} \nabla_{\nu'} - \frac{1}{2} g_{\mu \nu'} \nabla_\sigma \nabla^{\sigma'} + g_{\mu \nu'} \mu_\Phi^2 
    \right)\langle \Phi(x) \Phi(x') \rangle_\Psi, \label{def_tmiuniu}
\end{equation}
where the prim quantities refer to $x'$ and $g_{\mu \nu'}$ is defined by a transport along a geodesic curve connecting $x$ and $x'$. The equation \eqref{def_tmiuniu} is only formal because the right-hand side is plagued by the singularities that need to be regularized. Importantly, these singularities are universal for all Hadamard states given that they emerge from the local UV structure. 

In particular, the difference of two-point functions evaluated at two different states is smooth on any globally hyperbolic portion of a 
spacetime \cite{Radzikowski:1996pa}. Thus, while \eqref{def_tmiuniu} suffers from singularities, the difference between two energy-momentum tensors is perfectly well-defined. Hence, our goal will be to evaluate
\begin{equation}
    \langle T_{\mu \nu}(x) \rangle_\Psi - \langle T_{\mu \nu}(x) \rangle_\mathcal{C}
\end{equation}
close to the Cauchy horizon, where $\mathcal{C}$ is a Hadamard state in II and IV\footnote{One may worry about defining a state in the region IV for it contains a singularity. In practice, one can deform smoothly IV in a way that avoids the timelike singularity and does not affect our calculation.}. Since the state $\mathcal{C}$ is well-defined in IV and II, its energy-momentum tensor should be smooth at the Cauchy horizon. Thus, any singularity in $ \langle T_{\mu \nu}(x) \rangle_\Psi - \langle T_{\mu \nu}(x) \rangle_\mathcal{C}$ is due to $\langle T_{\mu \nu}(x) \rangle_\Psi$. Moreover, if we change the state $\Psi$ to any other Hadamard state on I and II, the answer may change only by a smooth function. Thus, a potential singular behavior of $\langle T_{\mu \nu}(x) \rangle_\Psi$ is a state-independent, universal feature. In particular, we want to show that in Eddington-Finkelstein coordinates at the Cauchy horizon, we have
\begin{equation}
        \langle T_{vv}(x) \rangle_\Psi - \langle T_{vv}(x) \rangle_\mathcal{C} \sim C \neq 0,
\end{equation}
where $C$ is a constant. By the standard chain rule, that would imply
\begin{equation}
    \langle T_{VV}(x) \rangle_\Psi \sim C V^{-2}
\end{equation}
in Kruskal-Szekeres coordinates and that is enough to prove the strong cosmic censorship. The authors in \cite{Hollands:2019whz} indeed showed that this is the case for the RN-dS black holes, as reviewed in Sec.~\ref{sec:review}. Although the same reasoning follows for any (stationary) Cauchy horizon, they found that the constant $C=0$ in the case of the rotating BTZ black hole, thus invalidating the desired conclusion. We will now show that, upon incorporating quantum corrections, $C$ does not vanish anymore.

Since that constant is state-independent, one may take $\Psi$ to be the Hartle-Hawking (HH) state. However, for technical reasons, it is easier to work with Boulware modes $\psi_{\omega, m}$ defined by their boundary data:
\begin{subequations}
    \begin{equation}
        \psi^I_{\omega m} = \frac{1}{2\pi \sqrt{2 |\omega|r_+}}e^{im \phi_+ - i \omega u} \quad \textrm{on\ } \mathcal{H}^-
    \end{equation}
    and
        \begin{equation}
        \psi^{II}_{\omega m} = \frac{1}{2\pi \sqrt{2 |\omega|r_+}}e^{im \phi_+ - i \omega u} \quad \textrm{on\ } \mathcal{H}^L
    \end{equation}
\end{subequations}
and vanishing at $\mathcal{H}^R$ and at infinity. In the above expressions, $\phi_+$ is the corotating angle and $u$ is retarded time. On the other hand, the comparison state is given by
\begin{equation}
    \psi^{out}_{\omega m} = \frac{1}{2\pi \sqrt{2 |\omega|r_-}}e^{im \phi_- - i \omega v} \quad \textrm{on\ } \mathcal{CH}^L.
\end{equation}
and zero on $\mathcal{CH}^R$.
 The key point is that the Hawking-Hartle state admits a very simple representation in terms of the Boulware modes:
\begin{align}
\begin{split}
    &\langle \lbrace \Phi(x_1), \Phi(x_2) \rbrace \rangle_{HH} = \int_{\mathbb{R}}d \omega \sum_{m \in \mathbb{Z}}  
    \frac{\textrm{sgn}( \omega)}{\sinh \left(
    \frac{\pi \omega}{\kappa_+}
    \right)}
    \mathrm{Re} \lbrace \psi^{up,I}_{\omega,m}(x_1) \overline{\psi}^{up,II}_{\omega,m}(x_2)\rbrace + \\
    & \int_{\mathbb{R}}d \omega \sum_{m \in \mathbb{Z}}   \frac{\textrm{sgn}( \omega)}{1- e^{-\frac{\pi \omega}{\kappa_+}}} \left(
\lbrace \psi^{up,I}_{\omega,m}(x_1) \overline{\psi}^{up,I}_{\omega,m}(x_2)\rbrace 
 + \lbrace \psi^{up,II}_{-\omega,m}(x_1) \overline{\psi}^{up,II}_{-\omega,m}(x_2)\rbrace  \right),
\end{split}
\end{align}
where $\lbrace \cdot, \cdot \rbrace$ is symmetrization and $\kappa_+$ is the event-horizon surface gravity. Some comments on technical subtleties are in order. Previously we encountered two coordinates on the quantum BTZ: local \eqref{qBTZ} and global \eqref{qBTZbar} ones. For our mode decomposition to be well-defined, $\phi$ should be in particular a periodic variable. Thus, all the quantities (for example $\omega, m, \kappa$) should be barred. From now on we will keep that distinction. Nevertheless, it is easier to work with local coordinates and change them at the very end. One just needs to remember that
\begin{subequations}
\begin{equation}
        m = \frac{1}{\Delta (1-\tilde{a}^2)} \left( 
\bar{m} - \bar{\omega} \tilde{a} \ell_3
    \right)
\end{equation}
and
\begin{equation}
    \omega = \frac{1}{\Delta (1-\tilde{a}^2)} \left( 
\bar{\omega} - \bar{m} \frac{\tilde{a}}{\ell_3}
    \right).
\end{equation}
\end{subequations}
Note that in particular, $m$ (in contrast to $\bar{m}$) is no longer an integer.

To properly calculate the energy-momentum tensor, we should understand better the solutions to the Klein-Gordon equation on our background. Following \cite{Hollands:2019whz}, let us introduce an auxiliary variable
\begin{equation}
    z = \frac{r^2 - r_2^2}{r_1^2 - r_2^2}.
\end{equation}
We may decompose solutions (in local coordinates) into Fourier modes $ e^{
i (m \phi - \omega t)} R_{\omega m}(z)$. For future reference, let us introduce the tortoise coordinate
\begin{equation}
    r^\star = \sum_i \frac{1}{2\kappa_i} \log |r-r_i|,
\end{equation}
where we sum over all roots of $H$. Here $\kappa_i = \frac{1}{2} H'(r_i)$ (so it differs from the standard definition by an absolute value). We will also need angular velocities:
\begin{equation}
    \Omega_i = \frac{\sqrt{r_1 r_2 r_3 r_4}}{r_i^2}.
\end{equation}
Let us introduce the parameters
\begin{align}
    \begin{split}
        \alpha &= \frac{1}{2}\left(
\Delta - i \frac{\omega - m \Omega_2}{|\kappa_2|} -i \frac{\omega - m \Omega_1}{\kappa_1}
        \right) \\
        \beta &= \frac{1}{2}\left(2-
\Delta - i \frac{\omega - m \Omega_2}{|\kappa_2|} -i \frac{\omega - m \Omega_1}{\kappa_1}
        \right) \\
        \gamma &= 1 - i \frac{\omega - m \Omega_2}{|\kappa_2|},
    \end{split}
\end{align}
where $\Delta = 1 + \sqrt{1+\mu_{\Phi}^2}$. 
The solutions with special behavior near the Cauchy horizon $r=r_2$ are given by
\begin{subequations}
    \begin{equation}
        R_{\omega m}^{out, -}(z) = z^{-\frac{1}{2}(1-\gamma)} (1-z)^{\frac{1}{2}(\alpha+\beta-\gamma)} f_{\omega m}^{out,-}(z)
    \end{equation}
    and
    \begin{equation}
                R_{\omega m}^{in, -}(z) = z^{\frac{1}{2}(1-\gamma)} (1-z)^{\frac{1}{2}(\alpha+\beta-\gamma)} f_{\omega m}^{in,-}(z).
    \end{equation}
\end{subequations}
The solutions with special behavior near the event horizon $r=r_1$ are given by
\begin{subequations}
    \begin{equation}
        R_{\omega m}^{out, +}(z) = z^{-\frac{1}{2}(1-\gamma)} |1-z|^{-\frac{1}{2}(\alpha+\beta-\gamma)} f_{\omega m}^{out,+}(z)
    \end{equation}
    and
    \begin{equation}
                R_{\omega m}^{in, +}(z) = z^{-\frac{1}{2}(1-\gamma)} |1-z|^{\frac{1}{2}(\alpha+\beta-\gamma)} f_{\omega m}^{in,+}(z).
    \end{equation}
\end{subequations}
Finally, the solution with the fast decay at infinity is given by
\begin{equation}
    R_{\omega m}^{0 \infty}(z) = z^{-\frac{1}{2}(2\alpha - \gamma + 1)} (z-1)^{\frac{1}{2}(\alpha+\beta-\gamma)} f_{\omega m}^{0 \infty}(z).
\end{equation}
All functions $f$ are chosen in such a way that they take value $1$ at an appropriate horizon (infinity). An attentive reader will notice that this is exactly the same asymptotic form as obtained in \cite{Hollands:2019whz}. However, functions $f$s satisfy much more complicated ODEs. In particular, they have four (not three) singular points and thus cannot be reduced to the hypergeometric equation. Nevertheless, we may introduce transmission-reflection coefficients:
\begin{subequations}
\begin{equation}
    R_{\omega m}^{out+} = \mathcal{A}_{\omega m} R_{\omega m}^{out-} + \mathcal{B}_{\omega m} R_{\omega m}^{in-}
\end{equation}
\begin{equation}
    R_{\omega m}^{in+} = \tilde{\mathcal{A}}_{\omega m} R_{\omega m}^{in-} + \tilde{\mathcal{B}}_{\omega m} R_{\omega m}^{out-}
\end{equation}
\begin{equation}
    \tilde{\mathcal{I}}_{\omega m} R^{0\inf}_{\omega m} = R_{\omega m}^{out+} + \tilde{\mathcal{R}}_{\omega m} R_{\omega m}^{in+}.
\end{equation}
\end{subequations}
However, we will have to find these coefficients numerically. 

Now, it is easy to see that 
\begin{equation}
    \phi^{in-}_{\omega m} = 2\pi \sqrt{2|\tilde\omega| r_2} \left( 
\frac{2 r_2}{r_1^2 - r_2^2}
    \right)^{-i \frac{\tilde\omega}{2\kappa_2}} (r_1 - r_2)^{i \frac{\tilde\omega}{2\kappa_1}} (r_2 - r_3)^{i \frac{\tilde\omega}{2\kappa_3}} (r_2 - r_4)^{i \frac{\tilde\omega}{2\kappa_4}} 
    \psi^{out}_{\tilde\omega, m},
\end{equation}
where $\tilde\omega = \omega - m \Omega_2$. 
Similarly we may write Boulware modes as
\begin{subequations}
\begin{equation}
    \psi^{out+}_{\omega m} +\tilde{\mathcal{R}}_{\omega m} \psi^{in+}_{\omega m} = 2\pi \sqrt{2|\tilde{\omega}| r_1} \left(\frac{2r_1}{r_1^2-r_2^2} \right)^{i \frac{\tilde{\omega}}{2\kappa_1}} (r_1-r_2)^{-i \frac{\tilde\omega}{2\kappa_2}} (r_1-r_3)^{-i \frac{\tilde\omega}{2\kappa_3}} (r_1-r_4)^{-i \frac{\tilde\omega}{2\kappa_4}} \psi^{I}_{\tilde\omega,m},
\end{equation}
    \begin{equation}
\psi^{out+}_{\omega m} = 2\pi \sqrt{2|\tilde{\omega}| r_1} \left(\frac{2r_1}{r_1^2-r_2^2} \right)^{i \frac{\tilde{\omega}}{2\kappa_1}} (r_1-r_2)^{-i \frac{\tilde\omega}{2\kappa_2}} (r_1-r_3)^{-i \frac{\tilde\omega}{2\kappa_3}} (r_1-r_4)^{-i \frac{\tilde\omega}{2\kappa_4}} \psi^{II}_{\tilde\omega,m},
    \end{equation}
\end{subequations}
where $\tilde{\omega}=\omega - m \Omega_+$. With that, we can finally calculate the two-point function. Close to the Cauchy horizon, it reads
\begin{align}
\begin{split}
    \langle \lbrace \Psi(x_1), &\Psi(x_2) \rbrace \rangle_{HH} = 
    \frac{r_2}{r_1} \sum_{\bar{m}\in\mathbb{Z}} \int_{\mathbb{R}} d\bar\omega \frac{|\tilde{\omega|}}{|\omega|} \\
    \times &\Big[ 
    \frac{\textrm{sgn}(\bar\omega)}{\sinh \left(
    \frac{\pi \bar\omega}{\bar\kappa_+}
    \right)}
    \mathrm{Re} \left(
    \left(\mathcal{B}_{\omega+m\Omega_1,m}
+ \tilde{\mathcal{R}}_{\omega+m\Omega_1,m} \tilde{\mathcal{A}}_{\omega+m\Omega_1,m}
    \right) \bar{\mathcal{B}}_{\omega+m\Omega_1,m}
    \lbrace \psi^{out}_{\tilde\omega,m}(x_1) \overline{\psi}^{out}_{\omega,m}(x_2)\rbrace \right)\\+
    & \frac{\textrm{sgn } (\bar\omega)}{1- e^{-\frac{\pi\bar\omega}{\bar\kappa_+}}} \Bigg(
\left|\mathcal{B}_{\omega+m\Omega_1,m}
+ \tilde{\mathcal{R}}_{\omega+m\Omega_1,m} \tilde{\mathcal{A}}_{\omega+m\Omega_1,m}
    \right|^2 \lbrace \psi^{out}_{\tilde\omega,m}(x_1) \overline{\psi}^{out}_{\tilde\omega,m}(x_2)\rbrace 
 \\+
 & |\mathcal{B}_{-\omega+m\Omega_1,m}|^2\lbrace \psi^{out}_{-\tilde\omega,m}(x_1) \overline{\psi}^{out}_{-\tilde\omega,m}(x_2)\rbrace \Bigg)
    \Big],
\end{split}
\end{align}
where $\tilde\omega = \omega + m \Omega_1 - m \Omega_2$ and $\omega, m$ are functions of $\bar\omega, \bar m$.


To obtain $\langle T_{vv} \rangle$, we need to calculate (regularized) $\langle \lbrace \partial_v\Psi(x_1), \partial_v \Psi(x_2) \rbrace \rangle_{HH}$ as $x_1 \to x_2 \to \mathcal{CH}$. This is immediate to obtain:
\begin{align}
\begin{split}
    \langle \lbrace \partial_v\Psi(x_1), \partial_v \Psi(x_2) \rbrace \rangle_{HH} &- \langle \lbrace \partial_v\Psi(x_1), \partial_v \Psi(x_2) \rbrace \rangle_{\mathcal{C}} = \\
    &-\frac{1}{8\pi^2 r_2^2} \sum_{\bar{m}} \int_{\mathbb{R}} d \bar{\omega} |\omega +m \Omega_1 - m\Omega_2| n_m(\omega),
\end{split}
\end{align}
where
\begin{align}
\begin{split}
    n_m(\omega) &= \frac{r_2}{r_1}  \frac{|\tilde{\omega|}}{|\omega|} 
    \times \Big[\frac{\textrm{sgn}(\bar\omega)}{\sinh \left(
    \frac{\pi \bar\omega}{\bar\kappa_+}
    \right)}
    \mathrm{Re} \left(
    \left(\mathcal{B}_{\omega+m\Omega_1,m}
+ \tilde{\mathcal{R}}_{\omega+m\Omega_1,m} \tilde{\mathcal{A}}_{\omega+m\Omega_1,m}
    \right) \bar{\mathcal{B}}_{\omega+m\Omega_1,m}\right) \\
    &+ \frac{\textrm{sgn } (\bar\omega)}{1- e^{-\frac{\pi\bar\omega}{\bar\kappa_+}}} \Bigg(
\left|\mathcal{B}_{\omega+m\Omega_1,m}
+ \tilde{\mathcal{R}}_{\omega+m\Omega_1,m} \tilde{\mathcal{A}}_{\omega+m\Omega_1,m}
    \right|^2 + |\mathcal{B}_{-\omega+m\Omega_1,m}|^2\Bigg)
    \Big] - \frac{1}{2} \left|\coth\frac{\pi \bar{\omega}}{\bar{\kappa_-}}\right|.
\end{split}
\end{align}
To calculate $n_m(\omega)$, we must find reflection-transmission coefficients. That can be done by solving a Klein-Gordon equation (that, after variable separation reduces to an ODE in $z$) with appropriate boundary conditions. To this end, we used the second-order Runge--Kutta method and then simply compared solutions at $z=0.4$ and $z=0.6$ (and analogously for solutions between the event horizon and infinity). The direct numerical calculation now shows that $n_m(\omega) \neq 0$ as long as quantum corrections are non-zero\footnote{Of course, for a fine-tuned choice of the black hole and field parameters, it may still happen that $C=0$. Such a configuration cannot be deemed as generic though and thus does not constitute a violation of the strong cosmic censorship. Moreover, we can expect that once higher quantum corrections are taken into account, the fine-tuning problem will go away.}. 

\section{Dilaton-gravity description}

\label{comments}
\label{sec:potential} 

Our dimensionally reduced metrics in two dimensions can be shown to correspond to a specific potential in the 2D dilaton-gravity description. The most famous example is the Jackiw-Teitelboim (JT) gravity \cite{Jackiw:1984je, Teitelboim:1983ux}, which,  among other things, encodes the near-horizon dynamics of many near-extremal black holes. This JT gravity will be ``modified'' here in the same sense that we have modified the BTZ solution by adding quantum corrections. In this sense, our metrics will be solutions of a generalized 2D dilaton-gravity theory with a specific potential \cite{Grumiller:2002nm}. In other words, here we will present a well-motivated example of a dilaton potential that gives as a solution a 2D black hole with a curvature singularity behind the horizon.

The action of a generalized dilaton-gravity takes the form\footnote{We are omitting boundary and topological terms since they do not alter the equations of motion.} 
\begin{equation}
    S = \frac{1}{16\pi G}\int d^2 x \sqrt{-g} \left( \phi R - U(\phi) \right),
\end{equation}
where $\phi$ represents the dilaton, and $U(\phi)$ is a potential whose exact form can be obtained on-shell by using the equations of motion\footnote{The rest of the equations of motion give $\nabla_\mu \nabla_\nu \phi + \frac{1}{2} U(\phi) g_{\mu \nu} = 0$ which are easily shown to be satisfied.}. From this action, we can reverse-engineer the potential that would give quantum BTZ as a solution. The variation over $\phi$ gives
\begin{equation}
    R = U'(\phi).
\end{equation}
From here we obtain for  $\phi = r$
\begin{equation}
    R = - H''(r) = -\frac{2}{L^2} + 2\frac{\ell \mu}{r^3} - \frac{6a^2}{r^4}
\end{equation}
Thus we see that
\begin{equation}
    U'(\phi) = -\frac{2}{L^2} + 2\frac{\ell \mu}{\phi^3}-\frac{6a^2}{\phi^4}
\end{equation}
and
\begin{equation}
    U(\phi) = -\frac{2}{L^2} \phi - \frac{\ell \mu}{\phi^2} + \frac{2a^2}{\phi^3}. \label{dila-pot}
\end{equation}
The first term corresponds to the 2D cosmological constant, and the last term is related to the spin of the black hole, as shown in App.~\ref{app:gauge-field}. We see that quantum corrections enter in a very simple way, and in particular, they do not depend on the spin of the black hole. Naturally, this result is just a generalization of \cite{Ach_carro_1993}, but it is worth noting that for $a = 0$, this potential corresponds to having a spacelike curvature singularity behind the black hole horizon. Consequently, we do not obtain a simple AdS$_2$ spacetime as a solution, but a spacetime modified by quantum corrections $\mu\ell$. The only regime in which this gives us the JT action is for large values of $\phi$ where quantum corrections fall off.

However, the 2D metric with a singularity behind the horizon is a result of dimensional reduction from a higher-dimensional solution. Therefore, we would like to obtain the solution for ``quantum JT'' from a dimensional reduction of a higher-dimensional action. The natural starting point is the brane action, defined already in \cite{Emparan_2020}, which consists of a 3D Einstein-Hilbert term, an infinite tower of higher-curvature corrections, but also of a CFT action that describes the quantum fields on the brane. The higher-curvature corrections have been consistently taken to zero in this paper: we are always working at the leading order in $\ell$, the parameter that sets the relevance of higher-curvature corrections\footnote{Naturally, they can be always included since their reduction is somewhat tedious but nevertheless straightforward.}. However, the brane quantum fields are harder to deal with due to their strong coupling. But as a toy model that captures well many of the features of quantum BTZ, we can look at the reduction with a conformally coupled massless free scalar.

\subsection{Reduction from 3D theory}
\label{sec:3D}

To obtain the action for dilaton-gravity, we start with the Einstein-Hilbert action in 3D, together with a conformally coupled scalar,
\begin{equation}
    I = I_{EH} + I_{\text{matter}},
\end{equation}
\begin{equation}
   I_{EH} = \frac{1}{16\pi \bar{G}} \int_{\mathcal{M}} d^{D}x \sqrt{\bar{g}} \left(\bar{R}- 2\lambda\right) + \frac{1}{8\pi \bar{G}} \int_{\partial{\mathcal{M}}} d^{D-1}x \sqrt{\bar{h}} \bar{K},
\end{equation}
\begin{equation}
      I_{\text{matter}} = \frac{1}{2} \int_{\mathcal{M}} d^{D}x \sqrt{\bar{g}} \left(\xi \bar{R} \psi^2 + (\partial \psi)^2 \right),
\end{equation}
where bars denote a 3D quantity, $\psi$ is the conformally coupled scalar, and $\xi = \frac{1}{8}$ for $D=3$, as shown in App.~\ref{app:ccs}\footnote{Note that one can obtain the JT action from a 3D theory in an alternative fashion, as outlined in \cite{Geng:2022slq, Geng:2022tfc, Aguilar-Gutierrez:2023tic}. In this case, the JT action is obtained from the higher-dimensional (3D) bulk, taking into account fluctuations transverse to the brane. On the other hand, we are looking directly at the brane theory, and performing a dimensional reduction to 2D.}. The dimensional reduction of this action is straightforward, but let us look at the full 3D solution first, which was found in \cite{Martinez:1996gn}. The solutions they found encompass the quantum BTZ case, but only for a special set of parameters. Namely, they find the blackening factor
\begin{equation}
    H(r) = \frac{r^2}{\ell_3^2} - a - \frac{b}{r}
\end{equation}
where $a$ and $b$ are related through other equations of motion,
\begin{equation}
    b = \frac{2a}{3} = \frac{2 B^2}{\ell_3^2}, \hspace{15pt} B \ge 0,
\end{equation}
and where the scalar field profile takes the form
\begin{equation}
    \psi = \frac{A}{\sqrt{r + B}}, \hspace{15pt} A = \sqrt{\frac{B}{\pi \bar{G}}}.
\end{equation}
We see that we can connect this solution only for one set of parameters $\mu$ and $\ell$, and consequently, obtaining back the standard BTZ solution is then impossible. These features arise because we have coupled a conformal and massless classical scalar field to gravity, and hence generated no new scales that one can use to tune to desired solutions. Of course, the quantum BTZ solution is obtained by coupling to a quantum scalar, which then introduces the Planck scale into the system. 

Nevertheless, they find that the Kretschmann scalar diverges at the inner horizon just like in the quantum BTZ case and that hence, there is a curvature singularity at $r=0$. If our goal is to simply obtain some dilaton potential for which the singularity persists at the 2D level, then it makes sense to look at the dimensional reduction of this system. Furthermore, this is a classical scalar that is coupled to gravity, so there are no suppressed additional modes when reducing the theory.

The details of the dimensional reduction can be found in App.~\ref{app:dim-red}, and here we only write the final answer\footnote{We redefined the scalar field in order to write the action in a more suitable form.},
\begin{equation}
    \frac{1}{16\pi G} \int_{\mathcal{M}} d^2x \sqrt{g} e^\phi \left(R - 2\lambda + (\partial \psi)^2 + \xi \psi^2 R - 2\xi \psi^2(\Box\phi + (\partial \phi)^2)\right) + \frac{1}{8\pi G} \int_{\partial{\mathcal{M}}} dx \sqrt{h} e^\phi K.
\end{equation}
Notice that we can rewrite the derivatives of the dilaton more conveniently, emphasizing their coupling with the scalar field,
\begin{equation}
\begin{split}
    \frac{1}{16\pi G} \int_{\mathcal{M}} d^2x \sqrt{g} e^\phi &\left(R(1 + \xi \psi^2) - 2\lambda + (\partial \psi)^2 + 2\psi (\partial_\mu \psi)(\partial^\mu \phi)\right)  \\
     &+ \frac{1}{8\pi G} \int_{\partial{\mathcal{M}}} dx \sqrt{h} e^\phi \left(K - \psi^2 n^\mu \partial_\mu \phi\right).
\end{split}
\end{equation}
Given that the 3D theory has as a solution a black hole with a singularity behind the horizon, the reduced theory will likewise have as a solution such a black hole. Integrating out the scalar field will then give a dilaton potential much like the one in equation \eqref{dila-pot} (with related parameters, as discussed above, and no angular momentum). Therefore, we have an example of a 3D theory that gives as a solution a singularity behind the horizon. It would be interesting to generalize this theory to include rotation, and look at it more closely from a CFT point of view, along the lines of \cite{Ghosh_2020}.

\subsection{Reduction from 4D theory}
\label{sec:4dto2d}

Even though in Sec.~\ref{sec:3D} we obtained a dilaton potential from a higher-dimensional theory, the solutions we get are very restricted and do not allow for parameter tuning. Moreover, there is no limit to the standard BTZ case. In this sense, a complete theory that has quantum BTZ as a solution would involve a quantum conformally coupled scalar at strong coupling. Given the difficulty of this problem, we can try to leverage the doubly holographic nature of this solution, and look at the higher-dimensional, purely gravitational parent theory in 4D. 

In other words, to properly account for quantum effects that stem from the brane CFT, we will perform the dimensional reduction directly in the 4D C-metric.  However, as we shall see, this is not a straightforward task, and in fact, a full reduction does not exist for this ansatz.

Let us focus on the non-rotating case, for which the C-metric takes the form
\begin{equation}
    ds^2 = \frac{\ell^2}{(\ell + xr)^2}\left[-H(r)dt^2 + \frac{dr^2}{H(r)} + r^2\left(\frac{dx^2}{G(x)} + G(x)d\varphi^2\right)\right],
\end{equation}
with
\begin{equation}
    H(r) = \frac{r^2}{\ell_3^2} + \kappa - \frac{\mu\ell}{r}, \hspace{15pt} G(x) = 1 - \kappa x^2 -\mu x^3.
\end{equation}
We will want to reduce the action of this theory on the sphere $(x,\varphi)$ by using the formula derived in App.~\ref{app:dim-red}, but there is a problem: the factor in front depends on one of the angular directions $x$. Therefore, we first have to Weyl rescale the metric, in order to obtain the correct form for the dimensional reduction procedure. In fact, this will be a problem for the rotating case as well, and the Weyl factor and the rescaling will be the same in both cases. The Weyl rescaling has a simple form,
\begin{equation}
    \bar{g}_{\mu\nu} = \omega^2 g_{\mu\nu}, \hspace{15pt} \omega(x,r) = \frac{\ell}{\ell + x r}.
\end{equation} 
The rescaling has been done in App.~\ref{app:weyl}, and here we only write the final results,
\begin{equation}
    \bar{R} = \omega^{-2} R - \frac{(D-1)(D-4)}{\omega^2} \left(\partial \omega\right)^2 - \frac{2(D-1)}{\omega} \Box \omega,
\end{equation}
\begin{equation}
    \bar{K} = \omega^{-1} K + \frac{D-1}{\omega^2} n^\alpha \partial_\alpha \omega.
\end{equation}
Now we have to replace the Ricci scalar and the extrinsic scalar with the dimensionally reduced quantities, derived in App.~\ref{app:dim-red}, and so our final scalars $R_F$ and $K_F$ will have the form
\begin{equation}
\begin{split}
    R_F &= \omega^{-2}\left(R^{(D-N)} - 2 N \Box \phi - N(N+1)(\partial \phi)^2 + e^{-2\phi} N(N-1)\right) \\
    &- \frac{(D-1)(D-4)}{\omega^2} \left(\partial \omega\right)^2 - \frac{2(D-1)}{\omega} \Box \omega,
\end{split}
\end{equation}
\begin{equation}
    K_F = \omega^{-1}\left(K^{(D-N-1)} + N n^\alpha \partial_\alpha \phi \right) + \frac{D-1}{\omega^2} n^\alpha \partial_\alpha \omega.
\end{equation}
In our case, $D=4$ and $N=2$, so we have
\begin{equation}
    \boxed{R_F = \omega^{-2}\left(R^{(2)} - 4\Box \phi - 6(\partial \phi)^2 + 2 e^{-2\phi}\right) - \frac{6}{\omega}\Box \omega} \label{boxR}
\end{equation}
and
\begin{equation}
    \boxed{K_F = \omega^{-1}\left(K^{(1)} + 2 n^\alpha \partial_\alpha \phi \right) + \frac{3}{\omega^2} n^\alpha \partial_\alpha \omega,} \label{boxK}
\end{equation}
where index $F$ stands for final. Now we should evaluate the metric density,
\begin{equation}
    \int d^{D-N}x\; d^Nx\; \omega^D \sqrt{-g_{D-N}} \sqrt{-g_{N}} = \int d^{D-N}x\; d^Nx\; \omega^D e^{N\phi} \sqrt{-g_{D-N}},
\end{equation}
where we see that there is an extra $\omega$ term, and so, our integral over the spherical components will not just give us $\Omega_N$ as in App.~\ref{app:dim-red}, but will be a function of $\omega$. 

One thing to have in mind is the fact that we have to use the full metric (non-reduced) for the derivatives of $\omega$, and so
\begin{equation}
    \Box \omega = \Box^{(x)} \omega + \Box^{(\theta)} \omega + N g^{\mu\nu} (\partial_\mu \omega) (\partial_\nu \phi),
\end{equation}
\begin{equation}
    (\partial \omega)^2 = (\partial^{(x)} \omega)^2 +  (\partial^{(\theta)} \omega)^2,
\end{equation}
where $x$ indicates the (d+1)-dimensional variables, $\theta$ indicates the $N$-dimensional ones, and the last term in the first expression comes from the fact that we have $\Gamma^\mu_{ab}$ (see App.~\ref{app:dim-red}). The indices $\mu$ and $\nu$ are over $x$ variables.

Our full action will then be
\begin{equation}
    \begin{split}
        \frac{1}{16\pi G_D} \int \sqrt{-g_{\mu\nu}} \sqrt{-\gamma_{ab}} \;\omega^4 e^{2\phi} \left(R_F - 2\lambda\right) + \frac{1}{8\pi G_D} \int \sqrt{-h_{\mu\nu}} \sqrt{-\gamma_{ab}}\; \omega^3 e^{2\phi} K_F, \label{actionred}
    \end{split}
\end{equation}
where we bear in mind that the integration is over all 4 variables. The tension of the brane is absorbed into $\lambda$. To see that we obtained the correct action, we have to check if this truncation is consistent---in other words, we have to insert the same ansatz in the D-dimensional action, and compare the equations of motion of the reduced and non-reduced action. If the equations of motion agree, then this is a consistent truncation. 
Nevertheless, this action is not reducible to 2D, for free $\omega$ and $\phi$, and so, we cannot make a meaningful comparison between the 3D and the 4D reduced theories. The best we can do is write
\begin{equation}
    I = I_{1} + I_{2},
\end{equation}
where 
\begin{equation}
\begin{split}
    I_1 &= \frac{1}{16\pi G_2} \int d^2x \sqrt{-g_{\mu\nu}} \chi_2(x) e^{2\phi} \left(R^{(2)} - 4\Box \phi - 6(\partial \phi)^2 + 2 e^{-2\phi} - \frac{2\chi_4(x)}{\chi_2(x)} \lambda \right) \\
    &+ \frac{1}{8\pi G_2} \int dx \sqrt{-h_{\mu\nu}} \chi_2(x) e^{2\phi} \left(K^{(1)} + 2 n^\alpha \partial_\alpha \phi \right),
\end{split}
\end{equation}
\begin{equation}
    I_2 = \frac{3}{8\pi G_2} \int \xi(x) = -\frac{3}{8\pi G_4} \int d^4x \sqrt{-g_4} e^{2\phi} \omega^3 \Box \omega + \frac{3}{8\pi G_4} \int d^3x \sqrt{-h_3} e^{2\phi} \omega n^\alpha \partial_\alpha \omega,
\end{equation}
where
\begin{equation}
    \chi_n(x) = \int d^2x \; \sqrt{-\gamma_{ab}} \;\omega^n,
\end{equation}
with $x$ being the non-spherical variables. Then, following \cite{Bueno:2022log}, the brane and bulk actions are connected as
\begin{equation}
    I_1 + I_2 = I_{\text{brane}} + I_{\text{CFT}},
\end{equation}
and from here, we obtain $I_{\text{CFT}}$ as the difference of the above terms. However, since we cannot reduce the action completely, we leave our findings in this form. 
We can further simplify $I_1$ by introducing a 2D Weyl rescaling, similar to what has been done in the App.~A of \cite{Svesko:2022txo}, but since we will not use this action further, we will not do so here (the procedure is identical to the aforementioned appendix).

One can, however, solve the on-shell action and obtain the relevant thermodynamic quantities. One can imagine also taking a limit in which the conformal factor disappears (for instance, when $\ell\to\infty$) and doing the reduction then, but in that case, the higher-curvature corrections on the brane become important and the three-dimensional character of the brane is lost. Finally, there may be a way to treat this action and look for the relevant modes without resorting to dimensional reduction, as suggested in \cite{mukund}.

In any case, the fact that we cannot fully reduce the 4D action tells us that the braneworld picture is obtained consistently only because certain degrees of freedom are frozen. This raises interesting questions about the consistency of braneworld theories. Nevertheless, we will see now that there exists a regime in which we can obtain a dimensional reduction to two dimensions.

\subsection*{The near-horizon ansatz}
\label{sec:nh reduction}

A reduction is possible with a different ansatz. Since the goal of this calculation is to obtain the action for the black hole, we can take a near-horizon limit of the 4D C-metric, $r = r_h + \xi$, with $\xi \ll 1$, which would zoom into the black hole, but also simplify the Weyl factor in front
\begin{equation}
    \omega(x,r) \to \omega(x) = \frac{\ell}{\ell + x r_h}.
\end{equation}
This way we obtain a Weyl factor that depends only on the spherical coordinate and not the radial one, and this simplification allows us to talk about dimensional reduction in a meaningful way\footnote{We thank Mukund Rangamani for discussions on this point.}. Namely, the equations \eqref{boxR} and \eqref{boxK} become
\begin{equation}
    R_F = \omega^{-2}\left(R^{(2)} - 4\Box \phi - 6(\partial \phi)^2 + 2 e^{-2\phi}\right) - \frac{6}{\omega}\Box \omega,
\end{equation}
\begin{equation}
    K_F = \omega^{-1}\left(K^{(1)} + 2 n^\alpha \partial_\alpha \phi \right),
\end{equation}
where now 
\begin{equation}
    \Box \omega = \Box^{(\theta)} \omega = \gamma^{ab}\nabla_a \nabla_b  \;\omega
\end{equation}
is a purely spherically dependent term. Our action \eqref{actionred} then becomes
\begin{equation}
\begin{split}
    I &= \frac{c_\omega}{16\pi G_D}\int d^2x \sqrt{-g_{\mu\nu}} \;e^{2\phi} \left(R^{(2)} - 4\Box \phi - 6(\partial \phi)^2 + 2 e^{-2\phi} \right) \\
    &+ I_\omega + \frac{c_\omega}{8\pi G_D} \int dx \sqrt{-h_{\mu\nu}} \;e^{2\phi} \left(K^{(1)} + 2 n^\alpha \partial_\alpha \phi\right),
\end{split}
\end{equation}
where
\begin{equation}
    c_\omega = \int d^2x \sqrt{-\gamma_{ab}} \;\omega^2,
\end{equation}
and
\begin{equation}
    I_\omega = -\frac{1}{16\pi G_D}\int d^4x \sqrt{-g_{\mu\nu}} \sqrt{-\gamma_{ab}} \;e^{2\phi} \omega^4 \left(\frac{6}{\omega}\Box \omega + 2\lambda \right) = -\frac{\hat{c}_\omega}{16\pi G_D} \int d^2x \sqrt{-g_{\mu\nu}}\; e^{2\phi},
\end{equation}
with
\begin{equation}
    \hat{c}_\omega = \int d^2x \sqrt{-\gamma_{ab}}\; \omega^4 \left(\frac{6}{\omega}\Box \omega + 2\lambda \right) = 2 c_\omega\lambda_\omega.
\end{equation}
The constant $c_\omega$ depends on the exact form of $\omega$ and on $\gamma_{ab}$, but at the end of the day, it is simply a constant with respect to the 2D spacetime we want to reduce to. We see that $I_\omega$ serves as a modified cosmological constant $\lambda_\omega$, so we can put it back in the action to obtain\footnote{We just have to keep in mind the labelling: $\lambda_\omega$ depends on $\omega$, and its contribution is really $\hat{c}_\omega$.}
\begin{equation}
    \begin{split}
    I &= \frac{c_\omega}{16\pi G_D}\int d^2x \sqrt{-g_{\mu\nu}} \;e^{2\phi} \left(R^{(2)} - 4\Box \phi - 6(\partial \phi)^2 + 2 e^{-2\phi} - 2\lambda_\omega \right) \\
    &+ \frac{c_\omega}{8\pi G_D} \int dx \sqrt{-h_{\mu\nu}} \;e^{2\phi} \left(K^{(1)} + 2 n^\alpha \partial_\alpha \phi\right).
\end{split}
\end{equation}
We see that we can dimensionally reduce our theory, but only if we restrict ourselves to the near-horizon limit of the higher-dimensional solution. Given that it is only an overall constant and the value of the cosmological constant that brings change with respect to the 3D reduction, we see that the near-horizon limit plays a similar role to the decoupling of the metric and the dilaton that was impossible with the full 4D ansatz. 

Of course, given that we are not dealing with a black hole that has an extremality parameter, the near-horizon limit simply reduces down to a Rindler description. Regardless, this simple exercise shows us that for the case of rotating black holes, we can expect a similar simplification to occur: the near-horizon limit makes the warping factor tractable, leading to a reducible action. In the case of a Kerr black hole in 4D, this simplification leads to the so-called near-horizon extremal Kerr (NHEK) geometry \cite{Bardeen:1999px}, which similarly has a warp factor that depends only on the angle\footnote{See recent work on obtaining logarithmic corrections to the near-extremal Kerr entropy \cite{kapec2023logarithmic, mukund}.}. We leave the calculation regarding the rotating C-metric for the follow-up work \cite{future}. 




\section{Discussion}
\label{sec:discussion}

In this paper, we have looked into the inner structure of black holes in three dimensions and their dimensionally reduced counterparts. Further, we have established the validity of these quantum black holes by checking they obey the relevant energy conditions, and we discussed some novel features.

The key focus of our work was the structure of singularities in quantum black holes: we established that the rotating quantum BTZ black hole really does obey strong cosmic censorship---both from the 2D and 3D perspective---with the only exception of infinitesimally near-extremal solutions, for which we know we would need to resort to a more detailed analysis anyway. One of the future directions we will undertake will be just that: how do the interior and the horizon of quantum black holes look once we include the effects coming from the Schwarzian? The horizon itself is not clearly defined in that case, so the calculations will be more subtle. It was also recently pointed out that near the extremality higher-curvature effects are additionally amplified \cite{Horowitz:2023xyl}. Thus, one may expect a plethora of quantum corrections in that regime.

Nevertheless, let us emphasize that the quantum rotating BTZ is a rare example of a geometry in which quantum corrections lead to qualitatively different results, even though we are in a low-curvature regime and far away from extremality. In particular, for a classical black hole and a massless scalar, Christodoulou’s version of the strong cosmic censorship is violated when $r_+<5 r_-$ \cite{Dias:2019ery}. It would be worthwhile to understand if there are other geometries for which quantum corrections may lift some accidental cancellations and thus lead to important new physics.

Additionally to checking the strong cosmic censorship, we have identified what type of a dilaton potential from the 2D perspective can give a proper curvature singularity behind the horizon. Previous studies have had smooth interiors, reflecting the simple symmetric nature of the AdS$_2$ spacetime, but we have seen that the addition of a large number of quantum fields significantly alters the interior structure even at the 2D level. Note also that the form of the dilaton potential \eqref{dila-pot} seems to be \textit{universal}, in the sense that it comes from the backreaction of matter fields onto the 3D black hole. Therefore, we have identified a physically well-motivated dilaton potential that leads to spacelike curvature singularities in 2D black holes. Obtaining the matrix model description of such singularities is a natural next step, left for a future study.

Another important next step will be to study the evaporation of these black holes, with the addition of the boundary graviton dynamics. This is especially interesting in the rotating case since one would like to see how the boundary graviton affects the evaporation rate of superradiant modes, which are the most relevant ones in the deep near-extremal case \cite{Brito:2015oca}. The use of the C-metric and the rotating quantum BTZ solution can be of help in this case since the rotating C-metric reduces to Kerr in the tensionless limit. In other words, the 3D brane black hole captures many essential features of the 4D Kerr black hole, while being analytically more tractable. One can then study the evaporation rate for the 3D black hole as a guiding toy model for the more realistic case. Finally, 
given that the rotating  (near)-extremal black hole is the least understood one from the perspective of its microstates \cite{Guica:2008mu, Ceplak:2023afb}, it is imperative to obtain the correct quantum picture for these black holes \cite{future}. 


\section*{Acknowledgements}

We thank Jan Boruch, Roberto Emparan, and Wayne Weng for their very useful comments on the draft of this paper. We thank Weam Abou Hamdan, Robie Hennigar, Stefan Hollands, Luca Iliesiu, Adam Levine, Sean McBride, Suvrat Raju, Mukund Rangamani, Harvey Reall, Mikel Sanchez Garitaonandia, Arvin Shahbazi-Moghaddam, Misha Usatyuk, and Jochen Zahn for useful discussions. MT is supported by the European Research Council (ERC) under the European Union’s Horizon 2020 research and innovation programme (grant agreement No 852386). 

\appendix

\section{Energy conditions}
\label{sec:energy}
It is worth checking if the quantum BTZ black holes satisfy some of the necessary energy conditions. Luckily, we find that the static case (also with higher-curvature corrections) satisfies the relevant energy condition (discussed below), while the rotating case has a restricted set of parameters, as expected. We also check the energy conditions for the de Sitter case and find that it is indeed allowed.

\subsection{The static case} The quantum-corrected geometry in \eqref{qBTZ} and \eqref{qBTZbar} is not a vacuum solution but is sourced by a stress tensor of the cutoff CFT. The leading order in the parameter $\ell$ of the quantum energy-momentum tensor that sources our geometry is
\begin{equation}
    \langle T^a_{\ b} \rangle_0 = \frac{\ell}{16\pi G_3} \frac{F(M)}{\overline{r}^3} \textrm{diag}(1,1,-2), \label{tmn}
\end{equation}
where the parameters are defined as in Sec.~\ref{sec:qBTZ}. Higher orders in $\ell$ indicate higher-curvature corrections, and we will consider them afterward, even though we can always work in a regime where such corrections are suppressed---this is the near-boundary limit. In fact, we will show that higher-curvature corrections do not change the validity obtained at the leading order.

Note that this stress tensor is traceless so the conformal symmetry is preserved at this order. However, one may be worried that it is not positive-definite. Indeed, let $\gamma$ be an affinely-parametrized null geodesic and
\begin{equation}
    u = u^t \partial_t + u^r \partial_r + u^\phi \partial_\phi
\end{equation}
a null vector tangent to $\gamma$, $u^\mu u_\mu = 0$. Then one can show from \eqref{qBTZbar} that \cite{Emparan_2020}
\begin{equation}
    \langle T_{ab} \rangle_0 u^a u^b = - \frac{3\ell}{16\pi G_3} \frac{F(M)}{\overline{r}} (u^\phi)^2 \le 0.
\end{equation}
Thus we see that the null energy condition is wildly violated. Even worse, if we integrate it along the geodesic
\begin{equation}
    \int_\gamma \langle T_{ab} \rangle_0 u^a u^b <0,
\end{equation}
we see that even the averaged null energy condition (ANEC) is violated unless $\gamma$ is non-rotating, $u^\phi = 0$. That would suggest that qBTZ  is not a sensible solution to the semiclassical Einstein equations. However, ANEC is supposed to hold only on geodesics that are complete and achronal \cite{Graham:2007va}. Achronality indicates that any two points on $\gamma$ cannot be connected by a timelike curve. In other words, it requires us to consider the \textit{fastest} null geodesic. 

Let us consider a null geodesic with non-zero $u^\phi$ and two points lying on it $p = (t_1, r_1, 0)$ and $q = (t_2, r_2,0)$. We can define two conserved quantities associated with two Killing vectors $\partial_t$ and $\partial_\phi$. These conserved quantities are the energy and the angular momentum,
\begin{equation}
    E = -K^\mu u_\mu = -u_t, \hspace{15pt} L = R^\mu u_\mu = u_\phi, \label{EL}
\end{equation}
where $K^\mu$ and $R^\mu$ are the timelike and angular Killing vectors, respectively,
\begin{equation}
    K^\mu = (1,0,0), \hspace{15pt} R^\mu = (0,0,1).
\end{equation}
With these conserved quantities, we can evaluate $u^r$, and determine the fastest geodesic between the points $p$ and $q$. Namely,
\begin{equation}
    t_2 - t_1 = \int_{r_1}^{r_2} \frac{E}{H \sqrt{E^2 - \frac{L^2}{\bar{r}^2}H}} d\bar{r}, \label{time}
\end{equation}
where $H$ is the blackening factor given by \eqref{qBTZbar}. If $r_1, r_2 > r_h$, we see that increasing the angular momentum leads to an increase in time that it takes to connect $p$ and $q$. In other words, the radial null geodesics will be the fastest ones, that is, achronal, and these clearly satisfy ANEC (and even NEC).

One might be more interested in the validity of black hole solutions in the dS$_3$ spacetime as discussed in \cite{Emparan:2022ijy}, since those do not exist classically. The stress tensor in that case is given by the same form as \eqref{tmn}, so the energy conditions will be satisfied in the same manner. However, we can additionally check if higher-curvature corrections spoil this solution. These are given by
\begin{equation}
\begin{split}
     \langle T^a_{\ b} \rangle_2 = \frac{\ell}{16\pi G_3} \frac{F(M)}{\overline{r}^3} \bigg(&-\frac{1}{2 R_3^2}\textrm{diag}(1,11,-10) - \frac{24 G_3 M}{\overline{r}^2 }\textrm{diag}(3,1,-4) \\
     &+ \frac{\ell F(M)}{2\overline{r}^3}\textrm{diag}(-29,-17,43)\bigg).
\end{split}
\end{equation}
The energy condition for radial null geodesics then gives
\begin{equation}
    \langle T^a_{\ b} \rangle u_a u^b = \left(T^r_r - T^t_t\right) g_{rr} (u^r)^2 > 0,
\end{equation}
as one can explicitly check, and so, our solution is valid even with higher-curvature terms taken into account.

\subsection{The rotating case} The situation is slightly more subtle in the case of the rotating qBTZ solution. Before we delve into the intricacies, we can first obtain the analogue of \eqref{time} for the rotating solution. Namely, the rotating qBTZ solution has the same Killing vectors as the non-rotating one, given by \eqref{EL}; the only difference will lie in the exact form of $u_t$ and $u_\phi$,
\begin{equation}
    u_t = g_{tt} u^t + g_{t\phi} u^\phi, \hspace{15pt} u_\phi = g_{\phi\phi} u^\phi + g_{t\phi} u^t.
\end{equation}
From the null condition $u^\mu u_\mu = 0$, we can then obtain $u^r$ and repeat the procedure from above. We then obtain 
\begin{equation}
    u^r = u^t \frac{dr}{dt} = \sqrt{\frac{E u^t - L u^\phi}{g_{rr}}},
\end{equation}
and using our conserved quantities, we get 
\begin{equation}
u^t = -\frac{1}{\text{det} g} \left(E g_{\phi\phi} + L g_{t\phi}\right), \hspace{15pt} u^\phi = \frac{1}{\text{det} g}\left(E g_{t\phi} + L g_{tt}\right),
\end{equation}

where
\begin{equation}
    \text{det} g = g_{tt} g_{\phi\phi} - g_{t\phi}^2.
\end{equation} 
We obtain 
\begin{equation}
    E u^t - L u^\phi = -\frac{1}{\text{det}g}\left(g_{\phi\phi} E^2 + 2g_{t\phi} E L + g_{tt} L^2\right),\label{expression}
\end{equation}
where for $g_{t\phi} = 0$, one obtains back the non-rotating result. The determinant factor is zero at the outer horizon since it is equal to the negative lapse function, and must be negative everywhere outside, so we can put $\text{det} g < 0$ since we will not evaluate the energy conditions inside the black hole.  

Unlike for the static case, it is not very obvious how to optimize the integrand with respect to the values of $E$ and $L$, so let us first bring it to the form that we require. Writing out the conserved expressions explicitly, we have
\begin{equation}
    \int dt = \int dr\;\sqrt{\frac{g_{rr} }{|\text{det} g|}} \frac{\alpha g_{\phi\phi} + g_{t\phi}}{\sqrt{\alpha^2 g_{\phi\phi} + 2 \alpha g_{t\phi} + g_{tt}}},
\end{equation}
where $\alpha \equiv E/L$. We can simplify this expression further by making a square to obtain
\begin{equation}
    \int dt = \int dr\; g(r) \left(1 - \frac{|\text{det} g|}{\beta^2}\right)^{-1/2},
\end{equation}
where 
\begin{equation}
    \beta = g_{\phi\phi}\left(\alpha + \frac{g_{t\phi}}{g_{\phi\phi}}\right), \hspace{15pt} g(r) = \sqrt{\frac{g_{rr} g_{\phi\phi}}{|\text{det} g|}}.
\end{equation}
One can easily check this expression gives the static case for $g_{t\phi} = 0$. And from here, it is clear that the fastest geodesics are those for which $\beta^2\to\infty$. Given that the metric coefficients are fixed, this implies $\alpha\to\pm\infty$, which is exactly the limit when $L\to 0$, regardless of its sign. This indicates that the fastest geodesics will be the ones with zero $L$, similar to the static case. Obviously, not every two points can be connected by $L=0$ geodesic. Nevertheless, at the risk of being not fully general, we will restrict to that case.

Having $L = 0$ implies $u^\phi = \frac{|g_{t\phi}|}{g_{\phi\phi}} u^t$, so from the null geodesic condition, we obtain
\begin{equation}
    (u^r)^2 = \frac{|\text{det} g|}{g_{rr} g_{\phi\phi}} (u^t)^2, \hspace{15pt} u^ru_r = - u^t u_t,
\end{equation}
so the null energy condition will then give
\begin{equation}
    \left(T_{tt} + T_{\phi\phi} \frac{g^2_{t\phi}}{g^2_{\phi\phi}} + 2 T_{t\phi} \frac{|g_{t\phi}|}{g_{\phi\phi}} + T_{rr} \frac{|\text{det}g|}{g_{rr} g_{\phi\phi}}\right)(u^t)^2 \ge 0.
\end{equation}
From \cite{Emparan_2020}, it is easier to evaluate the energy conditions in the form
\begin{equation}
    \left(T^t_{\ t} + T^t_{\ \phi} \frac{|g_{t\phi}|}{g_{\phi\phi}} - T^r_{\ r}\right)u^t u_t \ge 0, \label{rotstress}
\end{equation} 
and since $u^tu_t = -\frac{|\text{det} g|}{g_{\phi\phi}} (u^t)^2$, the expression in the parenthesis has to be smaller or equal to zero for the energy condition to be satisfied. The easiest case to check is the case of radial null geodesics which played a big role in the static case. However, first, we must change to an appropriate frame.

The local metric given by \eqref{rqBTZm} is not canonically normalized, and one would need to perform additional identifications in order to obtain an asymptotically AdS$_3$ solution. We need such a global form of the metric since we want to obtain geodesics that are complete and achronal, and both of these notions depend on the asymptotics and identifications. With these additional transformations, 
\begin{equation}
    t = \Delta\left(\bar{t} - \tilde{a} \ell_3 \bar{\phi}\right), \hspace{15pt} \phi = \Delta\left(\bar{\phi} - \frac{\tilde{a}}{\ell_3} \bar{t}\right), \hspace{15pt} r^2 = \frac{\bar{r}^2 - r_s^2}{(1 - \tilde{a}^2) \Delta^2}, \label{things}
\end{equation}
and,
\begin{equation}
    r_s = \frac{\ell_3 \tilde{a} \Delta}{x_1} \sqrt{2 - \kappa x_1^2}, \hspace{15pt} \tilde{a} = \frac{a x_1^2}{\ell_3},
\end{equation}
\noindent one arrives at a globally appropriate form of rotating qBTZ,
\begin{equation}
    ds^2 = -F_1(r)d\bar{t}^2 + \left(\bar{r}^2 + \frac{\ell_3^2 \ell \mu \Tilde{a}^2 \Delta^2}{r(\bar{r})}\right)d\bar{\phi}^2 - 8\mathcal{G}_3 J\left(1 + \frac{\ell}{x_1 r(\bar{r})}\right) d\bar{t} d\bar{\phi} + \frac{d\bar{r}^2}{F_2(r)},  \label{rqbtzg}
\end{equation}
where
\begin{equation}
    F_1(r) = \frac{\bar{r}^2}{\ell_3^2} - 8\mathcal{G}_3 M - \frac{\ell \mu \Delta^2}{r(\bar{r})},
\end{equation}

\begin{equation}
    F_2(r) = \frac{\bar{r}^2}{\ell_3^2} - 8\mathcal{G}_3 M + \frac{(4\mathcal{G}_3 J)^2}{\bar{r}^2} - \ell \mu (1 - \tilde{a}^2)^2 \Delta^4 \frac{r(\bar{r})}{\bar{r}^2},
\end{equation}
where we emphasized that in some places there is $r$ as a function of $\bar{r}$ given by \eqref{things}. The mass $M$ and the angular momentum $J$ are defined as
\begin{equation}
    M = -\frac{\kappa \Delta^2}{8\mathcal{G}_3} \left(1 + \tilde{a}^2 - \frac{4\tilde{a}^2}{\kappa x_1^2}\right), \hspace{15pt} J = \frac{\ell_3}{4\mathcal{G}_3} \tilde{a} \mu x_1 \Delta^2.
\end{equation}
The parameter $\tilde{a}$ introduces rotational effects and is connected to the standard parameter as $\tilde{a} = ax_1^2/\ell_3$, making $\tilde{a}$ dimensionless. We will restrict to $\tilde{a} \le 1$ so as to avoid regimes with closed timelike curves. The parameter $x_1$ has the same function as in the non-rotating case, and it simply gives us the portion of the higher-dimensional bulk that we are keeping, $0 \le x \le x_1$. Put differently, it sets the location of the brane. The renormalized Newton's constant is given by $\mathcal{G}_3 = G_3 \ell_4/\ell$, and we have 
\begin{equation}
    \mu = \frac{1 - \kappa x_1^2 + \tilde{a}^2}{x_1^3}.
\end{equation}
Other parameters are the same as in the non-rotating case. More details about this solution and how to obtain it can be found in \cite{Emparan_2020}. We kept the notations from their paper for easier comparison.

The stress tensor can now be obtained similarly to the non-rotating case, However, the expressions are more involved, as expected. Nevertheless, let us first check if the radial null geodesics satisfy the ANEC,
\begin{equation}
    8 \pi G_3 \langle T^{\bar{t}}_{\;\bar{t}} \rangle_0 = \frac{\ell \mu}{2 (1 - \tilde{a}^2) r^3} \left(1 + 2\tilde{a}^2 + \frac{3 \tilde{a}^2 \ell_3^2}{x_1^2 r^2}\right), \hspace{15pt} 8 \pi G_3 \langle T^{\bar{r}}_{\;\bar{r}} \rangle_0 = \frac{\ell \mu}{2 r^3}. \label{stresses}
\end{equation}
Now we can analyze the energy condition,
\begin{equation}
    \langle T_{ab} \rangle u^a u^b  = g_{\bar{t}\bar{t}} T^{\bar{t}}_{\; \bar{t}} (u^{\bar{t}})^2 + g_{\bar{r} \bar{r}} T^{\bar{r}}_{\; \bar{r}} (u^{\bar{r}})^2 = \left(T^{\bar{r}}_{\; \bar{r}} - T^{\bar{t}}_{\; \bar{t}}\right) g_{\bar{r} \bar{r}} (u^{\bar{r}})^2, 
\end{equation}
where we used the null geodesic condition. We see that in order for the ANEC to be satisfied, we must have $T^{\bar{r}}_{\; \bar{r}} - T^{\bar{t}}_{\; \bar{t}} \ge 0$, but from the forms given in \eqref{stresses}, we see that that is only possible in the non-rotating case $\tilde{a} = 0$, in which case we obtain exactly zero. For any $\tilde{a} \neq 0$, the difference is negative, and the ANEC is seemingly violated. To have a consistent solution, we must show that no radial null geodesics are complete. From the Penrose diagram, it is clear that no such radial geodesic can be complete, since they will all inevitably end up at the inner (Cauchy) horizon. However, there exists one geodesic for which this is not true: this is the horizon generator itself. By definition, the horizon is achronal, and no incompleteness is involved. Nevertheless, it is easy to see that the null geodesic tangent to the horizon is rotating: it must be proportional to the Killing vector field generating the horizon which is given by \cite{Emparan_2020}
\begin{equation}
    k = \partial_t+ \frac{a}{r_+^2} \partial_\phi = \frac{1 + \tilde{a}\ell_3}{\Delta(1-\tilde{a}^2)} \partial_{\bar{t}} + \frac{(x_1^2 r_+^2 + \ell_3^2) \tilde{a}}{\Delta (1-\tilde{a}^2) \ell_3 x_1^2 r_+^2} \partial{\bar{\phi}}, \label{kvhor}
\end{equation}
and so, we have no complete null radial geodesics. Therefore, there is no need to check the energy conditions in this case.

This indicates that one must allow for non-zero $u^\phi$, so we will need the extra information about another stress tensor component in order to solve for \eqref{rotstress}\footnote{If we had used the un-barred (local) coordinate system, then the $tt$ and $rr$ components of the stress tensor would be equivalent, and there would be no other relevant component, so the NEC would be trivially satisfied. However, this coordinate system implies a rotation of frames even in the $r\to\infty$ limit, so it is not a globally well-defined coordinate system. },
\begin{equation}
    8 \pi G_3 \langle T^{\bar{t}}_{\;\bar{\phi}} \rangle_0 = -\frac{3\ell \ell_3 \mu \tilde{a}}{2(1-\tilde{a}^2) r^3} \left(1 + \frac{\tilde{a}^2\ell_3^2}{x_1^2 r^2}\right).
\end{equation}
Since the $rr$-component is small compared to the $tt$-component, it is sufficient to compare just the $tt$ and $t\phi$ components. From here, one will obtain bounds on the various parameters of the rotating quantum BTZ black hole.

\section{Rescalings and reductions}
\label{app:appendix}

\subsection{Weyl rescaling}
\label{app:weyl}

We start with a metric in $D$ dimensions that is Weyl-rescaled,
\begin{equation}
    \bar{g}_{\mu\nu} = \omega^2 g_{\mu\nu}.
\end{equation}

The Christoffel symbols can then be written as 
\begin{equation}
    \bar{\Gamma}_{\mu\nu}^\rho = \Gamma_{\mu\nu}^\rho + \omega^{-1} g^{\rho\sigma}\left[(\partial_\mu \omega) g_{\sigma\nu} + (\partial_\nu \omega) g_{\sigma\mu} - (\partial_\sigma \omega) g_{\mu\nu}\right] = \Gamma_{\mu\nu}^\rho + \delta \Gamma_{\mu\nu}^\rho,
\end{equation}
so
\begin{equation}
    \delta \Gamma_{\mu\nu}^\rho = \frac{1}{\omega} \left((\partial_\mu \omega) \delta^\rho_\nu + (\partial_\nu \omega) \delta^\rho_\mu -(\partial_\sigma \omega) g^{\rho\sigma} g_{\mu\nu}\right), \hspace{15pt}  \delta \Gamma_{\rho\nu}^\rho = \frac{D}{\omega}\partial_\nu \omega.
\end{equation}

The Ricci tensor,
\begin{equation}
    \bar{R}_{\mu\nu} = \partial_\rho \bar{\Gamma}^\rho_{\mu\nu} - \partial_\mu \bar{\Gamma}^\rho_{\rho\nu} + \bar{\Gamma}^\rho_{\rho\sigma}\bar{\Gamma}^\sigma_{\mu\nu} - \bar{\Gamma}^\rho_{\mu\sigma}\bar{\Gamma}^\sigma_{\rho\nu},
\end{equation}
can be decomposed into
\begin{equation}
    \begin{split}
        \bar{R}_{\mu\nu} &= R_{\mu\nu} + \partial_\rho \delta \Gamma^\rho_{\mu\nu} - \partial_\mu \delta \Gamma^\rho_{\rho\nu} + \Gamma^\rho_{\rho\sigma}\delta \Gamma^\sigma_{\mu\nu} + \delta\Gamma^\rho_{\rho\sigma} \Gamma^\sigma_{\mu\nu} + \delta\Gamma^\rho_{\rho\sigma}\delta \Gamma^\sigma_{\mu\nu} \\ &- \Gamma^\rho_{\mu\sigma} \delta \Gamma^\sigma_{\rho\nu} - \delta\Gamma^\rho_{\mu\sigma} \Gamma^\sigma_{\rho\nu} - \delta\Gamma^\rho_{\mu\sigma} \delta \Gamma^\sigma_{\rho\nu}.
    \end{split}
\end{equation}
The first derivative terms give
\begin{equation}
\begin{split}
    \partial_\rho\delta \Gamma^\rho_{\mu\nu} - \partial_\mu \delta \Gamma^\rho_{\rho\nu} &= \frac{D-2}{\omega^2} (\partial_\mu \omega)(\partial_\nu \omega) + \frac{1}{\omega^2} (\partial \omega)^2 g_{\mu\nu} - \frac{D-2}{\omega} \partial_\mu \partial_\nu \omega \\
    & - \frac{1}{\omega}\left(\partial^2\omega g_{\mu\nu} + (\partial_\sigma \omega)(\partial_\rho g^{\rho\sigma})g_{\mu\nu} + (\partial_\sigma \omega) g^{\rho\sigma} (\partial_\rho g_{\mu\nu})\right),
\end{split}
\end{equation}
where we note
\begin{equation}
    (\partial_\sigma \omega)(\partial_\rho g^{\rho\sigma})g_{\mu\nu} = -g^{\rho\sigma} (\partial^\alpha\omega)(\partial_\rho g_{\alpha\sigma}) g_{\mu\nu}.
\end{equation}
The next 6 terms put together give
\begin{equation}
\begin{split}
    &\frac{D-2}{\omega} (\partial_\sigma \omega) \Gamma^\sigma_{\mu\nu} + \frac{D-2}{\omega^2}  (\partial_\mu \omega) (\partial_\nu\omega) - \frac{D-2}{\omega^2} (\partial \omega)^2 g_{\mu\nu} \\
    & + \frac{1}{\omega}\left( (\partial_\alpha\omega) g^{\alpha\sigma}\; \Gamma^\rho_{\sigma\nu} \;g_{\mu\rho} +  (\partial_\alpha \omega) g^{\alpha\sigma}\; \Gamma^\rho_{\mu\sigma} \;g_{\rho\nu} -  (\partial_\alpha \omega) g^{\alpha\sigma}\; \Gamma^\rho_{\rho\sigma} \;g_{\mu\nu}\right).
\end{split}
\end{equation}
The Ricci scalar is obtained simply,
\begin{equation}
    \bar{R} = \frac{1}{\omega^2} g^{\mu\nu} \bar{R}_{\mu\nu},
\end{equation}
and noting that 
\begin{equation}
    \Box \omega = g^{\mu\nu}\nabla_\mu \nabla_\nu \omega = \partial^2\omega - g^{\alpha\sigma} (\partial^\mu g_{\sigma\mu}) (\partial_\alpha \omega) + \frac{1}{2}g^{\alpha\sigma} g^{\mu\nu} (\partial_\sigma g_{\mu\nu})(\partial_\alpha \omega), 
\end{equation}
we obtain the desired result\footnote{In the case that we have $\bar{g}_{\mu\nu} = e^{2\Omega} g_{\mu\nu}$, the last part will add an additional $(\partial \Omega)^2$ term, and the full expression will be
\begin{equation}
    \bar{R} = e^{-2\Omega}\left( R - (D-1)(D-2) (\partial \Omega)^2 - 2(D-1) \Box \Omega \right).
\end{equation}},

\begin{equation}
   \boxed{ \bar{R} = \omega^{-2} R - \frac{(D-1)(D-4)}{\omega^2} (\partial \omega)^2 - \frac{2(D-1)}{\omega} \Box \omega.}
\end{equation}

The extrinsic curvature is obtained in a simple way. Noting that the indices are raised with $g_{\mu\nu}$, and that
\begin{equation}
    g_{\mu\nu} = h_{\mu\nu} - n_\mu n_\nu, \hspace{15pt} \bar{h}_{\mu\nu} = \omega^2 h_{\mu\nu},
\end{equation}
and 
\begin{equation}
    \bar{n}^\mu = \frac{n^\mu}{\omega}, \hspace{15pt} \bar{n}_\mu = \omega n_\mu, \hspace{15pt} n^\alpha n_\alpha = 1
\end{equation}
for a spacelike hypersurface, we obtain
\begin{equation}
\begin{split}
    \bar{K}_{\mu\nu} &= \frac{1}{2}\mathcal{L}_n \bar{h}_{\mu\nu} = \frac{1}{2}\left(\bar{n}^\alpha (\partial_\alpha \bar{h}_{\mu\nu}) + (\partial_\mu \bar{n}^\alpha) \bar{h}_{\alpha\nu} + (\partial_\nu \bar{n}^\alpha) \bar{h}_{\alpha\mu}\right) 
    = \omega K_{\mu\nu} +\\
    &+ \frac{1}{2}\left(2n^\alpha (\partial_\alpha \omega) h_{\mu\nu} - (\partial_\mu \omega) n^\alpha g_{\alpha\nu} + (\partial_\mu \omega) n^\alpha n_\alpha n_\nu - (\partial_\nu \omega)n^\alpha g_{\alpha\mu} + (\partial_\nu \omega)n^\alpha n_\alpha n_\mu\right) \\
    &= \omega K_{\mu\nu} + n^\alpha(\partial_\alpha \omega) h_{\mu\nu}.
\end{split}
\end{equation}
The extrinsic scalar is then easily obtained as
\begin{equation}
    \bar{K} = \frac{1}{\omega^2} g^{\mu\nu} \bar{K}_{\mu\nu},
\end{equation}
that is, since $g^{\mu\nu} h_{\mu\nu} = D-1$,
\begin{equation}
    \boxed{\bar{K} = \omega^{-1} K + \frac{D-1}{\omega^2} n^\alpha \partial_\alpha \omega.}
\end{equation}

\subsubsection*{Conformally coupled scalar}
\label{app:ccs}
In order to construct an action for a scalar field $\psi$ that is conformally coupled to the metric, we start with a general ansatz for the action
\begin{equation}
    \frac{1}{16\pi G} \int d^Dx \sqrt{g}\; \left( (\partial \psi)^2 + \xi R \psi^2\right),
\end{equation}
where we need to determine the value of $\xi$ for which this action is invariant under Weyl rescaling. Using the results outlined above for the Ricci scalar, and performing a rescaling of $\psi \to \omega^{-\frac{D-2}{2}} \psi$, we obtain 
\begin{equation}
    \xi = \frac{D-2}{4(D-1)}
\end{equation}
for a Weyl-invariant action.
\subsection{Dimensional reduction of diagonal metrics}
\label{app:dim-red}

We will first give a general prescription for performing dimensional reduction, and then we will focus on our case. Let us start with a metric
\begin{equation}
    d\bar{s}^2_{d+1+N} = ds^2_{d+1} + e^{2\phi(x)} d\Sigma_N, \hspace{15pt} d\Sigma_N = \gamma_{ab}dx^a dx^b, \hspace{15pt} g_{ab} = e^{2\phi(x)}\gamma_{ab},
\end{equation}
where $\phi(x)$ depends on the (d+1)-dimensional coordinates, and $\Sigma$ is a manifold on which we will perform the reduction, with $a, b = 1,\dots,N$.  Note that $D=d+1+N$. The barred quantities will be (d+1+N)-dimensional. Through direct calculation, one obtains 
\begin{equation}
    \bar{\Gamma}^\rho_{\mu\nu} = \Gamma^\rho_{\mu\nu}, \hspace{15pt}\bar{\Gamma}^a_{bc} =  \Gamma^a_{bc}, \hspace{15pt} \bar{\Gamma}^\mu_{ab} = -g^{\mu\nu} (\partial_\nu \phi) g_{ab}, \hspace{15pt}   \bar{\Gamma}^b_{a\mu} = \delta^b_a \partial_\mu \phi
\end{equation}
where the rest of the combinations are zero. For the Riemann curvature tensor, we will use the following notation
\begin{equation}
    \bar{R}^{\;\;\;\;\;\;\;\;D}_{ABC} = \partial_B \bar{\Gamma}^D_{AC} - \partial_A \bar{\Gamma}^D_{BC} + \bar{\Gamma}^E_{AC}\bar{\Gamma}^D_{EB} - \bar{\Gamma}^E_{BC}\bar{\Gamma}^D_{EA},
\end{equation}
where $A, B,...= 1,\dots,d+1+N$. From here, one obtains
\begin{equation}
    \bar{R}^{\;\;\;\;\;\; \sigma}_{\mu\nu\rho} = R^{\;\;\;\;\;\; \sigma}_{\mu\nu\rho},
\end{equation}
\begin{equation}
    \bar{R}^{\;\;\;\;\;d}_{abc} = R^{\;\;\;\;\;d}_{abc} + \bar{\Gamma}^\mu_{ac} \bar{\Gamma}^d_{\mu b} - \bar{\Gamma}^\mu_{bc} \bar{\Gamma}^d_{\mu a} = R^{\;\;\;\;\;d}_{abc} + (\partial \phi)^2\left(g_{bc} \delta^d_a - g_{ac}\delta^d_b\right),
\end{equation}
\begin{equation}
     \bar{R}^{\;\;\;\;\;\;b}_{\mu a \nu} = -\partial_\mu \bar{\Gamma}^b_{\nu a} + \bar{\Gamma}^\sigma_{\mu\nu}\bar{\Gamma}^b_{\sigma a} - \bar{\Gamma}^c_{a\nu}\bar{\Gamma}^b_{c\mu} = -\delta^b_a\left(\nabla_\mu \partial_\nu \phi + \partial_\mu \phi \partial_\nu \phi\right).
\end{equation}
From here, using
\begin{equation}
    \bar{R}_{AB} = \partial_E \bar{\Gamma}^E_{AB} - \partial_A \bar{\Gamma}^E_{BE} + \bar{\Gamma}^E_{ED} \bar{\Gamma}^D_{AB} -  \bar{\Gamma}^E_{AD} \bar{\Gamma}^D_{EB},
\end{equation}
we can obtain the Ricci tensors,
\begin{equation}
    \bar{R}_{\mu\nu} = R_{\mu\nu} - N\left(\nabla_\mu \partial_\nu \phi + \partial_\mu \phi \partial_\nu \phi\right),
\end{equation}
\begin{equation}
    \bar{R}_{ab} = R_{ab} - g_{ab}\left(N(\partial \phi)^2 + \Box \phi\right),
\end{equation}
where we used
\begin{equation}
\begin{split}
    \nabla_\mu \bar{\Gamma}^\mu_{ab} &= \partial_\mu \bar{\Gamma}^\mu_{ab} + \bar{\Gamma}^\mu_{A\mu}\bar{\Gamma}^{A}_{ab} - \bar{\Gamma}^{A}_{a\mu}\bar{\Gamma}^\mu_{Ab} - \bar{\Gamma}^A_{b\mu}\bar{\Gamma}^\mu_{aA} \\
    &= \partial_\mu \bar{\Gamma}^\mu_{ab} + \bar{\Gamma}^\mu_{\nu\mu}\bar{\Gamma}^{\nu}_{ab} - \bar{\Gamma}^{c}_{a\mu}\bar{\Gamma}^\mu_{cb} - \bar{\Gamma}^{c}_{b\mu}\bar{\Gamma}^\mu_{ac}.
\end{split}
\end{equation}
To obtain the Ricci scalars, we assume $\Sigma_N$ is a sphere $S^N$, which is a maximally symmetric spacetime with
\begin{equation}
    R_{abcd} = \frac{R_\Sigma}{N(N-1)}\left(g_{ac}g_{bd} - g_{bc}g_{ad}\right),
\end{equation}
and
\begin{equation}
    R_{ab} = \frac{R_\Sigma}{N} g_{ab}, \hspace{15pt} R_\Sigma = N(N-1).
\end{equation}
From here, we get
\begin{equation}
    \boxed{\Bar{R} = R - 2 N \Box \phi - N(N+1)(\partial \phi)^2 + e^{-2\phi} R_\Sigma.}
\end{equation}
Now we have to obtain the extrinsic curvature, whose definition is 
\begin{equation}
    \bar{K}_{AB} = \frac{1}{2} \mathcal{L}_n h_{AB} = \frac{1}{2}\left(n^C\partial_C h_{AB} + (\partial_A n^C)h_{CB} + (\partial_B n^C)h_{AC}\right) = \frac{1}{2}\mathcal{L}_n h_{\mu\nu} + \frac{1}{2}\mathcal{L}_n h_{ab},
\end{equation}
where $n^C$ is a unit normal vector of an embedded surface with an induced metric $h_{AB}$. Our hypersurface will be labeled by the Greek indices, therefore
\begin{equation}
    \frac{1}{2}\mathcal{L}_n h_{ab} = \frac{1}{2}n^\alpha \partial_\alpha h_{ab}, \hspace{15pt} \partial_\alpha\left(e^{2\phi(x)} \gamma_{ab}\right) = 2 h_{ab} \partial_\alpha \phi,
\end{equation}
from which we obtain
\begin{equation}
    \bar{K}_{AB} = K_{\mu\nu} + n^\alpha \partial_\alpha \phi h_{ab},
\end{equation}
and 
\begin{equation}
    \boxed{\bar{K} = K + N n^\alpha \partial_\alpha \phi.}
\end{equation}
The last piece that we will need is 
\begin{equation}
    \int_{\mathcal{\bar{M}}} d^{d+1+N}x \sqrt{\bar{g}} = \Omega_N \int_{\mathcal{M}} d^{d+1}x \sqrt{g} e^{N\phi},
\end{equation}
where $\Omega_N$ is the volume of the $N$-sphere, and $\mathcal{\bar{M}}$ the manifold on which we perform the reduction.

The Einstein-Hilbert action, together with a boundary term,
\begin{equation}
    \frac{1}{16\pi \bar{G}} \int_{\mathcal{\bar{M}}} d^{d+1+N}x \sqrt{\bar{g}} \;\left(\bar{R} - 2\lambda\right) + \frac{1}{8\pi \bar{G}} \int_{\partial\mathcal{\bar{M}}} d^{d+N}x \sqrt{\bar{h}} \;\bar{K},
\end{equation}
can now be written as 
\begin{equation}
\begin{split}
    \frac{1}{16 \pi G} \int_{\mathcal{M}} d^{d+1}x \sqrt{g} e^{N\phi} &\left(R - 2N\Box\phi - N(N+1)(\partial \phi)^2 + N(N-1)e^{-2\phi} -2\lambda\right) \\
    &+ \frac{1}{8\pi G} \int_{\partial\mathcal{M}} d^{d}x \sqrt{h}e^{N\phi}\left(K + Nn^\alpha \partial_\alpha \phi\right),
\end{split}
\end{equation}
where we set $G\equiv \frac{\bar{G}}{\Omega_N}$. This is the dimensionally reduced action on an $N$-sphere for a diagonal metric. We can further simplify this action by recalling Stokes' theorem, 
\begin{equation}
    \int_{\mathcal{M}} d^{d+1}x \sqrt{g}\; \nabla_\mu V^\mu = \int_{\partial\mathcal{M}} d^{d}x \sqrt{h}\; n_\mu V^\mu,
\end{equation}
for some generic vector $V^\mu$, to rewrite the term with $\Box \phi$ as
\begin{equation}
    -\frac{N}{8\pi G} \int_{\partial\mathcal{M}} \sqrt{h}\; e^{N\phi} n_\mu \nabla^\mu \phi + \frac{N^2}{8\pi G} \int_{\mathcal{M}} d^{d+1}x \sqrt{g}\; e^{N\phi} (\partial \phi)^2,
\end{equation}
where we can see that the first term cancels out exactly with the second term in the GHY contribution. The action is simplified even further for $N=1$,
\begin{equation}
    \frac{1}{16\pi G} \int_{\mathcal{M}} d^{d+1}x \sqrt{g}\; e^{\phi}\left(R - 2\lambda\right) + \frac{1}{8\pi G}\int_{\partial\mathcal{M}}  d^dx \sqrt{h} \; e^\phi K. 
\end{equation}

\subsection{Dimensional reduction of non-diagonal metrics}
\label{app:dim-redoff}

We start with a metric
\begin{equation}
    ds^2 = \bar{g}_{\mu\nu}dx^\mu dx^\nu = g_{ij} dx^i dx^j + k^2\left(d\varphi + A_i dx^i\right)^2,
\end{equation}
where none of the functions depend on $\varphi$, only on $x^i$. We want to reduce on $\varphi$, so we will need 
\begin{equation}
    \bar{g}_{\mu\nu} =
\begin{pmatrix}
k^2 & k^2 A_i \\
k^2 A_j & g_{ij} + k^2A_i A_j 
\end{pmatrix}, \hspace{15pt} \bar{g}^{\mu\nu} = \begin{pmatrix}
A_i A^i + k^{-2} & -A^j \\
-A^i & g^{ij}
\end{pmatrix},
\end{equation}
where 
\begin{equation}
    \text{det} \bar{g}_{\mu\nu} = k^2 g_{ij}. \label{determinant}
\end{equation}
The Christoffel symbols can be easily obtained:
\begin{equation}
    \bar{\Gamma}^0_{00} = k A^i (\partial_i k), \hspace{15pt} \bar{\Gamma}^i_{00} = -k g^{ij} (\partial_j k), \hspace{15pt} \bar{\Gamma}^0_{0i} = \frac{1}{k} (\partial_i k) + kA^j A_i (\partial_j k) - \frac{k^2}{2} A^j F_{ij},
\end{equation}
\begin{equation}
    \bar{\Gamma}^j_{i0} = \frac{k^2}{2} g^{jn} F_{in} - k A_i (\partial^j k), \hspace{15pt}  \bar{\Gamma}^n_{ij} = \Gamma^n_{ij} - k A_i A_j (\partial^n k) + k^2 g^{mn} A_{(i} F_{j)m},
\end{equation}
\begin{equation}
    \bar{\Gamma}^0_{ij} = \partial_{(i} A_{j)} + \frac{2}{k} A_{(i} (\partial_{j)} k) - A_n \Gamma^n_{ij} + k A_i A_j A^n (\partial_n k) - k^2 A^n A_{(i} F_{j)n},
\end{equation}
where
\begin{equation}
    A_{(i} B_{j)} \equiv \frac{1}{2}\left(A_i B_j + A_j B_i\right), \hspace{15pt} F_{ij} = \partial_i A_j - \partial_j A_i,
\end{equation}
and $\Gamma^n_{ij}$ is the Christoffel symbol w.r.t. $g_{ij}$. The ``$0$'' index refers to the $\varphi$ coordinate. The Ricci tensors are then given by 
\begin{equation}
    \bar{R}_{00} = -\frac{k^4}{4}F^{ij} F_{ij} - k (\partial^n k) \Gamma^j_{jn} - k g^{ij} (\partial_i k) (\partial_j k) - k (\partial_i g^{in}) (\partial_n k),
\end{equation}
\begin{equation}
\begin{split}
    \bar{R}_{0i} &= \frac{k^4}{4} F^{jn} F_{jn} A_i +  \frac{3k}{2} (\partial^l k) F_{il} + \frac{k^2}{2}\left((\partial_n g^{nl}) F_{il} + g^{nl} (\partial_n F_{il})\right) \\
    &- k A_i (\partial_i g^{in}) (\partial_n k) - k A_i g^{ij} \partial_i \partial_j k - k A_i (\partial^n k)  \Gamma^j_{jn} + \frac{k^2}{2} g^{jl} \left(F_{il} \Gamma^k_{kj} - F_{kl} \Gamma^k_{ij}\right),
\end{split}
\end{equation}
\begin{equation}
    \begin{split}
        \bar{R}_{ij} &= \frac{k^4}{4} F^{kn} F_{kn} A_i A_j - k A_i A_j (\partial_k g^{kn}) (\partial_n k) - k A_i A_j g^{mn} \partial_m \partial_n k \\
        & + \frac{3k}{2} A_i (\partial^n k) F_{jn} - k g^{lm} A_i A_j (\partial_m k) \Gamma^k_{kl} + \frac{3k}{2} A_j (\partial^n k) F_{in} - \frac{1}{k} \partial_j \partial_i k + \frac{1}{k} (\partial_n k) \Gamma^n_{ij} \\
        &+ \frac{k^2}{2} g^{kl} F_{ki} F_{jl} + k^2 (\partial_k g^{km}) A_{(i} F_{j)m} + k^2 g^{kn} (\partial_k F_{(in}) A_{j)} + k^2 g^{lm} \Gamma^k_{kl} A_{(i} F_{j)m} \\
        &- \frac{k^2}{2} g^{ln} A_i F_{kn} \Gamma^k_{jl} - \frac{k^2}{2} g^{km} A_j F_{lm} \Gamma^l_{ik} + \bar{R}_{ij}.
    \end{split}
\end{equation}
From here, we can obtain the Ricci scalar
\begin{equation}
    \bar{R} = \bar{g}^{00} \bar{R}_{00} + 2 \bar{g}^{0i} \bar{R}_{0i} + \bar{g}^{ij} \bar{R}_{ij},
\end{equation}
that is 
\begin{equation}
    \boxed{\bar{R} = R - \frac{k^2}{4} F^{ij} F_{ij} - \frac{2}{k} \Box k}
\end{equation}
where 
\begin{equation}
    \Box k = \nabla^i \nabla_i k = g^{ij} \left(\partial_i \partial_j k - \Gamma^n_{ij} (\partial_n k)\right).
\end{equation}

Writing in terms of $k = e^\phi$, we obtain
\begin{equation}
    \boxed{\bar{R} = R - \frac{e^{2\phi}}{4} F^2 - 2 \Box \phi - 2 (\partial \phi)^2}
\end{equation}
Setting $A_i = 0$ gives back the diagonal case for $N=1$.
For the extrinsic curvature, it is usual to choose a constant $x$ hypersurface, where $g_{\varphi x} = 0$. Therefore, we will have a normal vector $n^\alpha$, for which $h_{\alpha0} = 0$, that is $A_\alpha = 0$ (recall also that nothing depends on $\varphi$). This gives
\begin{equation}
    \bar{K}_{\mu\nu} = \frac{1}{2}\left(n^\alpha (\partial_\alpha h_{\mu\nu}) + (\partial_\mu n^\alpha) h_{\alpha\nu} + (\partial_\nu n^\alpha) h_{\mu \alpha} \right),
\end{equation}
that is
\begin{equation}
\begin{split}
    \bar{K}_{ij} &=  \frac{1}{2}\left(n^\alpha \left((\partial_\alpha h_{ij}) + 2 k A_i A_j (\partial_\alpha k) + k^2 (\partial_\alpha A_i) A_j + k^2 (\partial_\alpha A_j) A_i\right) + (\partial_i n^\alpha) h_{\alpha j} + (\partial_j n^\alpha) h_{i \alpha} \right)\\
    & = K_{ij} + \frac{1}{2} n^\alpha \left(2 k A_i A_j (\partial_\alpha k) + k^2 (\partial_\alpha A_i) A_j + k^2 (\partial_\alpha A_j) A_i\right),
\end{split}
\end{equation}
\begin{equation}
    \bar{K}_{i0} = \frac{1}{2} \left(n^\alpha (\partial_\alpha h_{i0}) + (\partial_i n^\alpha) h_{\alpha0} + (\partial_0 n^\alpha) h_{i\alpha}\right) = \frac{1}{2} n^\alpha (\partial_\alpha A_i) k^2 + k n^\alpha (\partial_\alpha k) A_i,
\end{equation}
\begin{equation}
    \bar{K}_{00} = \frac{1}{2} n^\alpha (\partial_\alpha h_{00}) = k n^\alpha (\partial_\alpha k).
\end{equation}
Put together, 
\begin{equation}
    \bar{K} = \bar{g}^{ij} \bar{K}_{ij} + 2 \bar{g}^{i0} \bar{K}_{i0} + \bar{g}^{00} \bar{K}_{00},
\end{equation}
we obtain
\begin{equation}
    \boxed{ \bar{K} = K + k^{-1} n^\alpha (\partial_\alpha k) }
\end{equation}
which is the same expression for the extrinsic curvature as in the diagonal case, 
\begin{equation}
    \boxed{\bar{K} = K +  n^\alpha (\partial_\alpha \phi) }
\end{equation}

\noindent The metric density is easily transformed \eqref{determinant},
\begin{equation}
    \int_{\mathcal{\bar{M}}} d^Dx \sqrt{\bar{g}} = 2\pi \int_{\mathcal{M}} d^{D-1} x \;k \sqrt{g},
\end{equation}
so we have an action
\begin{equation}
   \frac{1}{16 \pi G} \int_{\mathcal{M}} d^{D-1} x \;k \sqrt{g} \left(R - 2\lambda - \frac{k^2}{4} F^2 - \frac{2}{k} \Box k\right) + \frac{1}{8\pi G} \int_{\partial \mathcal{M}} d^{D-2}x \; k \sqrt{h} \left(K + \frac{1}{k} n^\alpha \partial_\alpha k\right), \label{rot-act}
\end{equation}
with $\bar{G} = \frac{G}{2\pi}$. Using Stokes' theorem, we see that the term with $\Box k$ exactly cancels the extra term in the boundary contribution, giving us the action 
\begin{equation}
     \frac{1}{16 \pi G} \int_{\mathcal{M}} d^{2} x \;k \sqrt{g} \left(R - 2\lambda - \frac{k^2}{4} F^2 \right) + \frac{1}{8\pi G} \int_{\partial \mathcal{M}} dx \; k \sqrt{h} K, \label{rot-act2}
\end{equation}
where we omitted the necessary counter-terms \cite{Emparan:1999pm}. We also set $D=3$ as it is the relevant case for us.

\subsection*{Integrating out the gauge field}
\label{app:gauge-field}

We see from the action \eqref{rot-act2} that the field strength tensor is a quadratic term, indicating that it should be possible to integrate it out. Indeed, this was done first in \cite{Ach_carro_1993}, and followed up with an alternative way, more suited for non-Abelian generalization, in \cite{Ghosh_2020}. Following \cite{Ach_carro_1993} for the 3D case, the field equation of motion gives
\begin{equation}
    \partial_\mu \left(F^{\mu\nu} k^3 \sqrt{g}\right) = 0 \hspace{5pt}\longrightarrow \hspace{5pt}F^{12} k^3 \sqrt{g} = \text{const},
\end{equation}
where $g = g_{tt} g_{rr}$. We can rewrite
\begin{equation}
    F^{tr} \sqrt{g}  = \frac{1}{\sqrt{g}} F_{tr} = \frac{1}{\sqrt{g}} \epsilon^{ij} \partial_i A_j,
\end{equation}
where 
\begin{equation}
    \epsilon^{ij}=
    \begin{cases}
      1, & \text{for (i,j) = (t,r)},  \\
      -1, & \text{for (i,j) = (r,t)}, \\
      0 & \text{otherwise}.
    \end{cases}
  \end{equation}

\noindent From here we obtain the field equation in \cite{Ach_carro_1993}
\begin{equation}
    \frac{k^3 \epsilon^{ij} \partial_i A_j}{\sqrt{g}} = \text{const} = \frac{k^3 \partial_r A_t}{\sqrt{g}}.
\end{equation}

The constant on the right-hand side is proportional to the spin $J$, as one can see through the Komar definition of the asymptotic charges \cite{carroll_2019},
\begin{equation}
    J = -\frac{1}{8\pi G} \int_{\partial \Sigma} dx \sqrt{\gamma} \; n_\mu \sigma_\nu \nabla^\mu R^\nu, 
\end{equation}
where $\partial \Sigma$ is a boundary metric defined by $\sqrt{\gamma} d\varphi = \sqrt{g_{\varphi \varphi}} d\varphi = k d\varphi$. The vectors $n_\mu$ and $\sigma_\nu$ are timelike and spacelike normal vectors, respectively,
\begin{equation}
    n_\mu = n_t = -\sqrt{g_{tt} + k^2 A_t^2}, \hspace{15pt} \sigma_\nu = \sigma_r = \sqrt{g_{rr}},
\end{equation}
and $R^\nu = R^\varphi = 1$ is the rotational Killing vector. Using the identity
\begin{equation}
    n_\mu \sigma_\nu \nabla^\mu R^\nu = R^\mu \sigma^\nu \nabla_\nu n_\mu  = K_{\varphi r} \sqrt{g^{rr}},
\end{equation}
where we used the Killing equation and the fact that $R^\mu$ and $n_\mu$ are normal, we obtain from
\begin{equation}
    K_{\varphi r} = \frac{k^2}{2} g^{tt} \partial_r A_t \sqrt{g_{tt} + k^2 A_t^2}
\end{equation}
that 
\begin{equation}
    J = -\frac{1}{4G} \frac{k^3 \partial_r A_t}{\sqrt{g}} \hspace{3pt} \Longrightarrow \hspace{3pt} \frac{k^3 \partial_r A_t}{\sqrt{g}} = -4 J G.
\end{equation}
Going back to our initial integral,
\begin{equation}
    I_{\text{Max}} = -\frac{1}{64\pi G} \int d^2x \; k^3 \sqrt{g} F^2 = -\frac{1}{16\pi G} \int d^2x \sqrt{g}\; \frac{(2 J G)^2}{k^3},
\end{equation}
we finally have an action dependent only on the dilaton and the metric,
\begin{equation}
 \frac{1}{16 \pi G} \int_{\mathcal{M}} d^{2} x \; \sqrt{g} \left(k R - U(k)\right) + \frac{1}{8\pi G} \int_{\partial \mathcal{M}} dx \; k \sqrt{h} K,
\end{equation}
where $U(k) = k^{-3} (2J G)^2 - 2\lambda k$.

\bibliography{bibi}

\providecommand{\href}[2]{#2}\begingroup\raggedright\begin{thebibliography}{10}

\bibitem{Dixon:1985jw}
L.~J. Dixon, J.~A. Harvey, C.~Vafa, and E.~Witten, ``{Strings on Orbifolds},''
  \href{http://dx.doi.org/10.1016/0550-3213(85)90593-0}{{\em Nucl. Phys. B}
  {\bfseries 261} (1985) 678--686}.

\bibitem{Aspinwall:1993yb}
P.~S. Aspinwall, B.~R. Greene, and D.~R. Morrison, ``{Multiple mirror manifolds
  and topology change in string theory},''
  \href{http://dx.doi.org/10.1016/0370-2693(93)91428-P}{{\em Phys. Lett. B}
  {\bfseries 303} (1993) 249--259},
  \href{http://arxiv.org/abs/hep-th/9301043}{{\ttfamily arXiv:hep-th/9301043}}.

\bibitem{Witten:1993yc}
E.~Witten, ``{Phases of N=2 theories in two-dimensions},''
  \href{http://dx.doi.org/10.1016/0550-3213(93)90033-L}{{\em Nucl. Phys. B}
  {\bfseries 403} (1993) 159--222},
  \href{http://arxiv.org/abs/hep-th/9301042}{{\ttfamily arXiv:hep-th/9301042}}.

\bibitem{Strominger:1995cz}
A.~Strominger, ``{Massless black holes and conifolds in string theory},''
  \href{http://dx.doi.org/10.1016/0550-3213(95)00287-3}{{\em Nucl. Phys. B}
  {\bfseries 451} (1995) 96--108},
  \href{http://arxiv.org/abs/hep-th/9504090}{{\ttfamily arXiv:hep-th/9504090}}.

\bibitem{Horowitz:1989bv}
G.~T. Horowitz and A.~R. Steif, ``{Space-Time Singularities in String
  Theory},'' \href{http://dx.doi.org/10.1103/PhysRevLett.64.260}{{\em Phys.
  Rev. Lett.} {\bfseries 64} (1990) 260}.

\bibitem{Horowitz:1990ap}
G.~T. Horowitz and A.~R. Steif, ``{Singular string solutions with nonsingular
  initial data},'' \href{http://dx.doi.org/10.1016/0370-2693(91)91214-G}{{\em
  Phys. Lett. B} {\bfseries 258} (1991) 91--96}.

\bibitem{Belinsky:1970ew}
V.~A. Belinsky, I.~M. Khalatnikov, and E.~M. Lifshitz, ``{Oscillatory approach
  to a singular point in the relativistic cosmology},''
  \href{http://dx.doi.org/10.1080/00018737000101171}{{\em Adv. Phys.}
  {\bfseries 19} (1970) 525--573}.

\bibitem{Belinski:2017fas}
V.~Belinski and M.~Henneaux,
  \href{http://dx.doi.org/10.1017/9781107239333}{{\em {The Cosmological
  Singularity}}}.
\newblock Cambridge Monogr.Math.Phys. Cambridge Univ. Pr., Cambridge, 2017.

\bibitem{Witten:2022xxp}
E.~Witten, ``{A Note On The Canonical Formalism for Gravity},''
  \href{http://arxiv.org/abs/2212.08270}{{\ttfamily arXiv:2212.08270
  [hep-th]}}.

\bibitem{Ori:1995nj}
A.~Ori and E.~E. Flanagan, ``{How generic are null space-time
  singularities?},'' \href{http://dx.doi.org/10.1103/PhysRevD.53.R1754}{{\em
  Phys. Rev. D} {\bfseries 53} (1996) 1754--1758},
  \href{http://arxiv.org/abs/gr-qc/9508066}{{\ttfamily arXiv:gr-qc/9508066}}.

\bibitem{Luk:2013cqa}
J.~Luk, ``{Weak null singularities in general relativity},''
  \href{http://dx.doi.org/10.1090/jams/888}{{\em J. Am. Math. Soc.} {\bfseries
  31} no.~1, (2018) 1--63}, \href{http://arxiv.org/abs/1311.4970}{{\ttfamily
  arXiv:1311.4970 [gr-qc]}}.

\bibitem{Dafermos:2017dbw}
M.~Dafermos and J.~Luk, ``{The interior of dynamical vacuum black holes I: The
  $C^0$-stability of the Kerr Cauchy horizon},''
  \href{http://arxiv.org/abs/1710.01722}{{\ttfamily arXiv:1710.01722 [gr-qc]}}.

\bibitem{Emparan:2021yon}
R.~Emparan and M.~Toma\v{s}evi\'c, ``{Quantum backreaction on chronology
  horizons},'' \href{http://dx.doi.org/10.1007/JHEP02(2022)182}{{\em JHEP}
  {\bfseries 02} (2022) 182}, \href{http://arxiv.org/abs/2109.03611}{{\ttfamily
  arXiv:2109.03611 [hep-th]}}.

\bibitem{Bousso:2022tdb}
R.~Bousso and A.~Shahbazi-Moghaddam, ``{Quantum singularities},''
  \href{http://dx.doi.org/10.1103/PhysRevD.107.066002}{{\em Phys. Rev. D}
  {\bfseries 107} no.~6, (2023) 066002},
  \href{http://arxiv.org/abs/2206.07001}{{\ttfamily arXiv:2206.07001
  [hep-th]}}.

\bibitem{PhysRevD.44.314}
E.~Witten, ``String theory and black holes,''
  \href{http://dx.doi.org/10.1103/PhysRevD.44.314}{{\em Phys. Rev. D}
  {\bfseries 44} (Jul, 1991) 314--324}.
  \url{https://link.aps.org/doi/10.1103/PhysRevD.44.314}.

\bibitem{Mertens:2022irh}
T.~G. Mertens and G.~J. Turiaci, ``{Solvable models of quantum black holes: a
  review on Jackiw\textendash{}Teitelboim gravity},''
  \href{http://dx.doi.org/10.1007/s41114-023-00046-1}{{\em Living Rev. Rel.}
  {\bfseries 26} no.~1, (2023) 4},
  \href{http://arxiv.org/abs/2210.10846}{{\ttfamily arXiv:2210.10846
  [hep-th]}}.

\bibitem{Stanford:2017thb}
D.~Stanford and E.~Witten, ``{Fermionic Localization of the Schwarzian
  Theory},'' \href{http://dx.doi.org/10.1007/JHEP10(2017)008}{{\em JHEP}
  {\bfseries 10} (2017) 008}, \href{http://arxiv.org/abs/1703.04612}{{\ttfamily
  arXiv:1703.04612 [hep-th]}}.

\bibitem{Jafferis:2022wez}
D.~L. Jafferis, D.~K. Kolchmeyer, B.~Mukhametzhanov, and J.~Sonner, ``{JT
  gravity with matter, generalized ETH, and Random Matrices},''
  \href{http://arxiv.org/abs/2209.02131}{{\ttfamily arXiv:2209.02131
  [hep-th]}}.

\bibitem{Grumiller:2002nm}
D.~Grumiller, W.~Kummer, and D.~V. Vassilevich, ``{Dilaton gravity in
  two-dimensions},''
  \href{http://dx.doi.org/10.1016/S0370-1573(02)00267-3}{{\em Phys. Rept.}
  {\bfseries 369} (2002) 327--430},
  \href{http://arxiv.org/abs/hep-th/0204253}{{\ttfamily arXiv:hep-th/0204253}}.

\bibitem{Grumiller:2021cwg}
D.~Grumiller, R.~Ruzziconi, and C.~Zwikel, ``{Generalized dilaton gravity in
  2d},'' \href{http://dx.doi.org/10.21468/SciPostPhys.12.1.032}{{\em SciPost
  Phys.} {\bfseries 12} no.~1, (2022) 032},
  \href{http://arxiv.org/abs/2109.03266}{{\ttfamily arXiv:2109.03266
  [hep-th]}}.

\bibitem{Witten:2020wvy}
E.~Witten, ``{Matrix Models and Deformations of JT Gravity},''
  \href{http://dx.doi.org/10.1098/rspa.2020.0582}{{\em Proc. Roy. Soc. Lond. A}
  {\bfseries 476} no.~2244, (2020) 20200582},
  \href{http://arxiv.org/abs/2006.13414}{{\ttfamily arXiv:2006.13414
  [hep-th]}}.

\bibitem{Maxfield:2020ale}
H.~Maxfield and G.~J. Turiaci, ``{The path integral of 3D gravity near
  extremality; or, JT gravity with defects as a matrix integral},''
  \href{http://dx.doi.org/10.1007/JHEP01(2021)118}{{\em JHEP} {\bfseries 01}
  (2021) 118}, \href{http://arxiv.org/abs/2006.11317}{{\ttfamily
  arXiv:2006.11317 [hep-th]}}.

\bibitem{Turiaci:2020fjj}
G.~J. Turiaci, M.~Usatyuk, and W.~W. Weng, ``{2D dilaton-gravity, deformations
  of the minimal string, and matrix models},''
  \href{http://dx.doi.org/10.1088/1361-6382/ac25df}{{\em Class. Quant. Grav.}
  {\bfseries 38} no.~20, (2021) 204001},
  \href{http://arxiv.org/abs/2011.06038}{{\ttfamily arXiv:2011.06038
  [hep-th]}}.

\bibitem{Blommaert:2022lbh}
A.~Blommaert, J.~Kruthoff, and S.~Yao, ``{An integrable road to a perturbative
  plateau},'' \href{http://dx.doi.org/10.1007/JHEP04(2023)048}{{\em JHEP}
  {\bfseries 04} (2023) 048}, \href{http://arxiv.org/abs/2208.13795}{{\ttfamily
  arXiv:2208.13795 [hep-th]}}.

\bibitem{Maldacena:1997re}
J.~M. Maldacena, ``{The Large N limit of superconformal field theories and
  supergravity},'' \href{http://dx.doi.org/10.4310/ATMP.1998.v2.n2.a1}{{\em
  Adv. Theor. Math. Phys.} {\bfseries 2} (1998) 231--252},
  \href{http://arxiv.org/abs/hep-th/9711200}{{\ttfamily arXiv:hep-th/9711200}}.

\bibitem{Fidkowski:2003nf}
L.~Fidkowski, V.~Hubeny, M.~Kleban, and S.~Shenker, ``{The Black hole
  singularity in AdS / CFT},''
  \href{http://dx.doi.org/10.1088/1126-6708/2004/02/014}{{\em JHEP} {\bfseries
  02} (2004) 014}, \href{http://arxiv.org/abs/hep-th/0306170}{{\ttfamily
  arXiv:hep-th/0306170}}.

\bibitem{Festuccia:2008zx}
G.~Festuccia and H.~Liu, ``{A Bohr-Sommerfeld quantization formula for
  quasinormal frequencies of AdS black holes},''
  \href{http://dx.doi.org/10.1166/asl.2009.1029}{{\em Adv. Sci. Lett.}
  {\bfseries 2} (2009) 221--235},
  \href{http://arxiv.org/abs/0811.1033}{{\ttfamily arXiv:0811.1033 [gr-qc]}}.

\bibitem{Dodelson:2023vrw}
M.~Dodelson, C.~Iossa, R.~Karlsson, and A.~Zhiboedov, ``{A thermal product
  formula},'' \href{http://arxiv.org/abs/2304.12339}{{\ttfamily
  arXiv:2304.12339 [hep-th]}}.

\bibitem{Horowitz:2023ury}
G.~T. Horowitz, H.~Leung, L.~Queimada, and Y.~Zhao, ``{Boundary signature of
  singularity in the presence of a shock wave},''
  \href{http://arxiv.org/abs/2310.03076}{{\ttfamily arXiv:2310.03076
  [hep-th]}}.

\bibitem{Frolov:1991nv}
V.~P. Frolov, ``{Vacuum polarization in a locally static multiply connected
  space-time and a time machine problem},''
  \href{http://dx.doi.org/10.1103/PhysRevD.43.3878}{{\em Phys. Rev. D}
  {\bfseries 43} (1991) 3878--3894}.

\bibitem{Kay:1996hj}
B.~S. Kay, M.~J. Radzikowski, and R.~M. Wald, ``{Quantum field theory on
  space-times with a compactly generated Cauchy horizon},''
  \href{http://dx.doi.org/10.1007/s002200050042}{{\em Commun. Math. Phys.}
  {\bfseries 183} (1997) 533--556},
  \href{http://arxiv.org/abs/gr-qc/9603012}{{\ttfamily arXiv:gr-qc/9603012}}.

\bibitem{Hollands:2019whz}
S.~Hollands, R.~M. Wald, and J.~Zahn, ``{Quantum instability of the Cauchy
  horizon in Reissner\textendash{}Nordstr\"om\textendash{}deSitter
  spacetime},'' \href{http://dx.doi.org/10.1088/1361-6382/ab8052}{{\em Class.
  Quant. Grav.} {\bfseries 37} no.~11, (2020) 115009},
  \href{http://arxiv.org/abs/1912.06047}{{\ttfamily arXiv:1912.06047 [gr-qc]}}.

\bibitem{Simpson:1973ua}
M.~Simpson and R.~Penrose, ``{Internal instability in a Reissner-Nordstrom
  black hole},'' \href{http://dx.doi.org/10.1007/BF00792069}{{\em Int. J.
  Theor. Phys.} {\bfseries 7} (1973) 183--197}.

\bibitem{Poisson:1989zz}
E.~Poisson and W.~Israel, ``{Inner-horizon instability and mass inflation in
  black holes},'' \href{http://dx.doi.org/10.1103/PhysRevLett.63.1663}{{\em
  Phys. Rev. Lett.} {\bfseries 63} (1989) 1663--1666}.

\bibitem{Ori:1991zz}
A.~Ori, ``{Inner structure of a charged black hole: An exact mass-inflation
  solution},'' \href{http://dx.doi.org/10.1103/PhysRevLett.67.789}{{\em Phys.
  Rev. Lett.} {\bfseries 67} (1991) 789--792}.

\bibitem{Hintz:2016gwb}
P.~Hintz and A.~Vasy, ``{The global non-linear stability of the Kerr-de Sitter
  family of black holes},'' \href{http://arxiv.org/abs/1606.04014}{{\ttfamily
  arXiv:1606.04014 [math.DG]}}.

\bibitem{Hintz:2016jak}
P.~Hintz, ``{Non-linear stability of the Kerr-Newman-de Sitter family of
  charged black holes},'' \href{http://arxiv.org/abs/1612.04489}{{\ttfamily
  arXiv:1612.04489 [math.AP]}}.

\bibitem{Dias:2018etb}
O.~J.~C. Dias, H.~S. Reall, and J.~E. Santos, ``{Strong cosmic censorship:
  taking the rough with the smooth},''
  \href{http://dx.doi.org/10.1007/JHEP10(2018)001}{{\em JHEP} {\bfseries 10}
  (2018) 001}, \href{http://arxiv.org/abs/1808.02895}{{\ttfamily
  arXiv:1808.02895 [gr-qc]}}.

\bibitem{Luna:2019olw}
R.~Luna, M.~Zilh\~ao, V.~Cardoso, J.~a.~L. Costa, and J.~Nat\'ario, ``{Strong
  cosmic censorship: The nonlinear story},''
  \href{http://dx.doi.org/10.1103/PhysRevD.99.064014}{{\em Phys. Rev. D}
  {\bfseries 99} no.~6, (2019) 064014},
  \href{http://arxiv.org/abs/1810.00886}{{\ttfamily arXiv:1810.00886 [gr-qc]}}.
  [Addendum: Phys.Rev.D 103, 104043 (2021)].

\bibitem{Kunduri:2007vf}
H.~K. Kunduri, J.~Lucietti, and H.~S. Reall, ``{Near-horizon symmetries of
  extremal black holes},''
  \href{http://dx.doi.org/10.1088/0264-9381/24/16/012}{{\em Class. Quant.
  Grav.} {\bfseries 24} (2007) 4169--4190},
  \href{http://arxiv.org/abs/0705.4214}{{\ttfamily arXiv:0705.4214 [hep-th]}}.

\bibitem{Kunduri:2013gce}
H.~K. Kunduri and J.~Lucietti, ``{Classification of near-horizon geometries of
  extremal black holes},'' \href{http://dx.doi.org/10.12942/lrr-2013-8}{{\em
  Living Rev. Rel.} {\bfseries 16} (2013) 8},
  \href{http://arxiv.org/abs/1306.2517}{{\ttfamily arXiv:1306.2517 [hep-th]}}.

\bibitem{Holzegel:2011uu}
G.~Holzegel and J.~Smulevici, ``{Decay properties of Klein-Gordon fields on
  Kerr-AdS spacetimes},'' \href{http://dx.doi.org/10.1002/cpa.21470}{{\em
  Commun. Pure Appl. Math.} {\bfseries 66} (2013) 1751--1802},
  \href{http://arxiv.org/abs/1110.6794}{{\ttfamily arXiv:1110.6794 [gr-qc]}}.

\bibitem{Hartnoll:2020rwq}
S.~A. Hartnoll, G.~T. Horowitz, J.~Kruthoff, and J.~E. Santos, ``{Gravitational
  duals to the grand canonical ensemble abhor Cauchy horizons},''
  \href{http://dx.doi.org/10.1007/JHEP10(2020)102}{{\em JHEP} {\bfseries 10}
  (2020) 102}, \href{http://arxiv.org/abs/2006.10056}{{\ttfamily
  arXiv:2006.10056 [hep-th]}}.

\bibitem{Randall:1999vf}
L.~Randall and R.~Sundrum, ``{An Alternative to compactification},''
  \href{http://dx.doi.org/10.1103/PhysRevLett.83.4690}{{\em Phys. Rev. Lett.}
  {\bfseries 83} (1999) 4690--4693},
  \href{http://arxiv.org/abs/hep-th/9906064}{{\ttfamily arXiv:hep-th/9906064}}.

\bibitem{Bueno:2022log}
P.~Bueno, R.~Emparan, and Q.~Llorens, ``{Higher-curvature gravities from
  braneworlds and the holographic c-theorem},''
  \href{http://dx.doi.org/10.1103/PhysRevD.106.044012}{{\em Phys. Rev. D}
  {\bfseries 106} no.~4, (2022) 044012},
  \href{http://arxiv.org/abs/2204.13421}{{\ttfamily arXiv:2204.13421
  [hep-th]}}.

\bibitem{Emparan:2022ijy}
R.~Emparan, J.~F. Pedraza, A.~Svesko, M.~Toma\v{s}evi\'c, and M.~R. Visser,
  ``{Black holes in dS$_{3}$},''
  \href{http://dx.doi.org/10.1007/JHEP11(2022)073}{{\em JHEP} {\bfseries 11}
  (2022) 073}, \href{http://arxiv.org/abs/2207.03302}{{\ttfamily
  arXiv:2207.03302 [hep-th]}}.

\bibitem{Emparan_2020}
R.~Emparan, A.~M. Frassino, and B.~Way, ``Quantum {BTZ} black hole,''
  \href{http://dx.doi.org/10.1007/jhep11(2020)137}{{\em Journal of High Energy
  Physics} {\bfseries 2020} no.~11, (Nov, 2020) }.
  \url{https://doi.org/10.1007%2Fjhep11%282020%29137}.

\bibitem{Emparan:2020rnp}
R.~Emparan and M.~Toma\v{s}evi\'c, ``{Strong cosmic censorship in the BTZ black
  hole},'' \href{http://dx.doi.org/10.1007/JHEP06(2020)038}{{\em JHEP}
  {\bfseries 06} (2020) 038}, \href{http://arxiv.org/abs/2002.02083}{{\ttfamily
  arXiv:2002.02083 [hep-th]}}.

\bibitem{Emparan:1999fd}
R.~Emparan, G.~T. Horowitz, and R.~C. Myers, ``{Exact description of black
  holes on branes. 2. Comparison with BTZ black holes and black strings},''
  \href{http://dx.doi.org/10.1088/1126-6708/2000/01/021}{{\em JHEP} {\bfseries
  01} (2000) 021}, \href{http://arxiv.org/abs/hep-th/9912135}{{\ttfamily
  arXiv:hep-th/9912135}}.

\bibitem{Emparan:2002px}
R.~Emparan, A.~Fabbri, and N.~Kaloper, ``{Quantum black holes as holograms in
  AdS brane worlds},''
  \href{http://dx.doi.org/10.1088/1126-6708/2002/08/043}{{\em JHEP} {\bfseries
  08} (2002) 043}, \href{http://arxiv.org/abs/hep-th/0206155}{{\ttfamily
  arXiv:hep-th/0206155}}.

\bibitem{Emparan:2021xdy}
R.~Emparan and M.~Toma\v{s}evi\'c, ``{Holography of time machines},''
  \href{http://dx.doi.org/10.1007/JHEP03(2022)212}{{\em JHEP} {\bfseries 03}
  (2022) 212}, \href{http://arxiv.org/abs/2107.14200}{{\ttfamily
  arXiv:2107.14200 [hep-th]}}.

\bibitem{Tomasevic:2023ojy}
M.~Toma\v{s}evi\'c, ``{On the Inaccessibility of Time Machines},''
  \href{http://dx.doi.org/10.3390/universe9040159}{{\em Universe} {\bfseries 9}
  no.~4, (2023) 159}.

\bibitem{Frassino:2022zaz}
A.~M. Frassino, J.~F. Pedraza, A.~Svesko, and M.~R. Visser,
  ``{Higher-Dimensional Origin of Extended Black Hole Thermodynamics},''
  \href{http://dx.doi.org/10.1103/PhysRevLett.130.161501}{{\em Phys. Rev.
  Lett.} {\bfseries 130} no.~16, (2023) 161501},
  \href{http://arxiv.org/abs/2212.14055}{{\ttfamily arXiv:2212.14055
  [hep-th]}}.

\bibitem{Emparan:2021hyr}
R.~Emparan, A.~M. Frassino, M.~Sasieta, and M.~Toma\v{s}evi\'c, ``{Holographic
  complexity of quantum black holes},''
  \href{http://dx.doi.org/10.1007/JHEP02(2022)204}{{\em JHEP} {\bfseries 02}
  (2022) 204}, \href{http://arxiv.org/abs/2112.04860}{{\ttfamily
  arXiv:2112.04860 [hep-th]}}.

\bibitem{Panella:2023lsi}
E.~Panella and A.~Svesko, ``{Quantum Kerr-de Sitter black holes in three
  dimensions},'' \href{http://arxiv.org/abs/2303.08845}{{\ttfamily
  arXiv:2303.08845 [hep-th]}}.

\bibitem{Martinez:1996gn}
C.~Martinez and J.~Zanelli, ``{Conformally dressed black hole in
  (2+1)-dimensions},'' \href{http://dx.doi.org/10.1103/PhysRevD.54.3830}{{\em
  Phys. Rev. D} {\bfseries 54} (1996) 3830--3833},
  \href{http://arxiv.org/abs/gr-qc/9604021}{{\ttfamily arXiv:gr-qc/9604021}}.

\bibitem{Cisterna:2023qhh}
A.~Cisterna, F.~Diaz, R.~B. Mann, and J.~Oliva, ``{Exploring Accelerating Hairy
  Black Holes in $2+1$ Dimensions: The Asymptotically Locally Anti-de Sitter
  Class and its Holography},''
  \href{http://arxiv.org/abs/2309.05559}{{\ttfamily arXiv:2309.05559
  [hep-th]}}.

\bibitem{Engelhardt:2015gla}
N.~Engelhardt and G.~T. Horowitz, ``{Holographic Consequences of a No
  Transmission Principle},''
  \href{http://dx.doi.org/10.1103/PhysRevD.93.026005}{{\em Phys. Rev. D}
  {\bfseries 93} no.~2, (2016) 026005},
  \href{http://arxiv.org/abs/1509.07509}{{\ttfamily arXiv:1509.07509
  [hep-th]}}.

\bibitem{Birrell78}
N.~Birrell and P.~Davies, ``{On falling through a black hole into another
  universe},'' \href{http://dx.doi.org/https://doi.org/10.1038/272035a0}{{\em
  Nature} {\bfseries 272} (1978) 35}.

\bibitem{Shrivastava:2020xmw}
P.~Shrivastava, ``{Quantum aspects of charged black holes in de-Sitter
  space},'' \href{http://arxiv.org/abs/2009.03261}{{\ttfamily arXiv:2009.03261
  [hep-th]}}.

\bibitem{Hollands:2020qpe}
S.~Hollands, C.~Klein, and J.~Zahn, ``{Quantum stress tensor at the Cauchy
  horizon of the Reissner\textendash{}Nordstr\"om\textendash{}de Sitter
  spacetime},'' \href{http://dx.doi.org/10.1103/PhysRevD.102.085004}{{\em Phys.
  Rev. D} {\bfseries 102} no.~8, (2020) 085004},
  \href{http://arxiv.org/abs/2006.10991}{{\ttfamily arXiv:2006.10991 [gr-qc]}}.

\bibitem{Bhattacharjee:2020nul}
S.~Bhattacharjee, S.~Sarkar, and A.~Bhattacharyya, ``{Scalar perturbations of
  black holes in Jackiw-Teitelboim gravity},''
  \href{http://dx.doi.org/10.1103/PhysRevD.103.024008}{{\em Phys. Rev. D}
  {\bfseries 103} no.~2, (2021) 024008},
  \href{http://arxiv.org/abs/2011.08179}{{\ttfamily arXiv:2011.08179 [gr-qc]}}.

\bibitem{Moitra:2020ojo}
U.~Moitra, ``{Strong Cosmic Censorship in Two Dimensions},''
  \href{http://dx.doi.org/10.1103/PhysRevD.103.L081502}{{\em Phys. Rev. D}
  {\bfseries 103} no.~8, (2021) L081502},
  \href{http://arxiv.org/abs/2011.03499}{{\ttfamily arXiv:2011.03499
  [hep-th]}}.

\bibitem{Papadodimas:2019msp}
K.~Papadodimas, S.~Raju, and P.~Shrivastava, ``{A simple quantum test for
  smooth horizons},'' \href{http://dx.doi.org/10.1007/JHEP12(2020)003}{{\em
  JHEP} {\bfseries 12} (2020) 003},
  \href{http://arxiv.org/abs/1910.02992}{{\ttfamily arXiv:1910.02992
  [hep-th]}}.

\bibitem{Dias:2019ery}
O.~J.~C. Dias, H.~S. Reall, and J.~E. Santos, ``{The BTZ black hole violates
  strong cosmic censorship},''
  \href{http://dx.doi.org/10.1007/JHEP12(2019)097}{{\em JHEP} {\bfseries 12}
  (2019) 097}, \href{http://arxiv.org/abs/1906.08265}{{\ttfamily
  arXiv:1906.08265 [hep-th]}}.

\bibitem{Ghosh_2020}
A.~Ghosh, H.~Maxfield, and G.~J. Turiaci, ``A universal Schwarzian sector in
  two-dimensional conformal field theories,''
  \href{http://dx.doi.org/10.1007/jhep05(2020)104}{{\em Journal of High Energy
  Physics} {\bfseries 2020} no.~5, (May, 2020) }.
  \url{https://doi.org/10.1007%2Fjhep05%282020%29104}.

\bibitem{Iliesiu:2020qvm}
L.~V. Iliesiu and G.~J. Turiaci, ``{The statistical mechanics of near-extremal
  black holes},'' \href{http://dx.doi.org/10.1007/JHEP05(2021)145}{{\em JHEP}
  {\bfseries 05} (2021) 145}, \href{http://arxiv.org/abs/2003.02860}{{\ttfamily
  arXiv:2003.02860 [hep-th]}}.

\bibitem{Christodoulou:2008nj}
D.~Christodoulou, \href{http://dx.doi.org/10.1142/9789814374552_0002}{``{The
  Formation of Black Holes in General Relativity},''} in {\em {12th Marcel
  Grossmann Meeting on General Relativity}}, pp.~24--34.
\newblock 5, 2008.
\newblock \href{http://arxiv.org/abs/0805.3880}{{\ttfamily arXiv:0805.3880
  [gr-qc]}}.

\bibitem{Radzikowski:1996pa}
M.~J. Radzikowski, ``{Micro-local approach to the Hadamard condition in quantum
  field theory on curved space-time},''
  \href{http://dx.doi.org/10.1007/BF02100096}{{\em Commun. Math. Phys.}
  {\bfseries 179} (1996) 529--553}.

\bibitem{Jackiw:1984je}
R.~Jackiw, ``{Lower Dimensional Gravity},''
  \href{http://dx.doi.org/10.1016/0550-3213(85)90448-1}{{\em Nucl. Phys. B}
  {\bfseries 252} (1985) 343--356}.

\bibitem{Teitelboim:1983ux}
C.~Teitelboim, ``{Gravitation and Hamiltonian Structure in Two Space-Time
  Dimensions},'' \href{http://dx.doi.org/10.1016/0370-2693(83)90012-6}{{\em
  Phys. Lett. B} {\bfseries 126} (1983) 41--45}.

\bibitem{Ach_carro_1993}
A.~Ach\'{u}carro and M.~E. Ortiz, ``Relating black holes in two and three
  dimensions,'' \href{http://dx.doi.org/10.1103/physrevd.48.3600}{{\em Physical
  Review D} {\bfseries 48} no.~8, (Oct, 1993) 3600--3605}.
  \url{https://doi.org/10.1103%2Fphysrevd.48.3600}.

\bibitem{Geng:2022slq}
H.~Geng, A.~Karch, C.~Perez-Pardavila, S.~Raju, L.~Randall, M.~Riojas, and
  S.~Shashi, ``{Jackiw-Teitelboim Gravity from the Karch-Randall Braneworld},''
  \href{http://dx.doi.org/10.1103/PhysRevLett.129.231601}{{\em Phys. Rev.
  Lett.} {\bfseries 129} no.~23, (2022) 231601},
  \href{http://arxiv.org/abs/2206.04695}{{\ttfamily arXiv:2206.04695
  [hep-th]}}.

\bibitem{Geng:2022tfc}
H.~Geng, ``{Aspects of AdS$_{2}$ quantum gravity and the Karch-Randall
  braneworld},'' \href{http://dx.doi.org/10.1007/JHEP09(2022)024}{{\em JHEP}
  {\bfseries 09} (2022) 024}, \href{http://arxiv.org/abs/2206.11277}{{\ttfamily
  arXiv:2206.11277 [hep-th]}}.

\bibitem{Aguilar-Gutierrez:2023tic}
S.~E. Aguilar-Gutierrez, A.~K. Patra, and J.~F. Pedraza, ``{Entangled universes
  in dS wedge holography},'' \href{http://arxiv.org/abs/2308.05666}{{\ttfamily
  arXiv:2308.05666 [hep-th]}}.

\bibitem{Svesko:2022txo}
A.~Svesko, E.~Verheijden, E.~P. Verlinde, and M.~R. Visser, ``{Quasi-local
  energy and microcanonical entropy in two-dimensional nearly de Sitter
  gravity},'' \href{http://dx.doi.org/10.1007/JHEP08(2022)075}{{\em JHEP}
  {\bfseries 08} (2022) 075}, \href{http://arxiv.org/abs/2203.00700}{{\ttfamily
  arXiv:2203.00700 [hep-th]}}.

\bibitem{mukund}
I.~Raki\'{c}, M.~Rangamani, and G.~J. Turiaci, ``Near-extremal Kerr and its
  entropy,''
  \href{http://arxiv.org/abs/https://www2.yukawa.kyoto-u.ac.jp/~qimg2023/presentation\_files/Rangamani\_Mukund\_09\_14\_.pdf}{{\ttfamily
  https://www2.yukawa.kyoto-u.ac.jp/~qimg2023/presentation\_files/Rangamani\_Mukund\_09\_14\_.pdf}}.

\bibitem{Bardeen:1999px}
J.~M. Bardeen and G.~T. Horowitz, ``{The Extreme Kerr throat geometry: A Vacuum
  analog of AdS(2) x S**2},''
  \href{http://dx.doi.org/10.1103/PhysRevD.60.104030}{{\em Phys. Rev. D}
  {\bfseries 60} (1999) 104030},
  \href{http://arxiv.org/abs/hep-th/9905099}{{\ttfamily arXiv:hep-th/9905099}}.

\bibitem{kapec2023logarithmic}
D.~Kapec, A.~Sheta, A.~Strominger, and C.~Toldo, ``{Logarithmic Corrections to
  Kerr Thermodynamics},'' \href{http://arxiv.org/abs/2310.00848}{{\ttfamily
  arXiv:2310.00848 [hep-th]}}.

\bibitem{future}
W.~Abou~Hamdan, N.~\v{C}eplak, M.~Kolanowski, and M.~Toma\v{s}evi\'{c},
  ``{Spontaneous superradiance of near-extremal black holes},'' {\em in
  progress} (2023) .

\bibitem{Horowitz:2023xyl}
G.~T. Horowitz, M.~Kolanowski, G.~N. Remmen, and J.~E. Santos, ``{Extremal Kerr
  Black Holes as Amplifiers of New Physics},''
  \href{http://dx.doi.org/10.1103/PhysRevLett.131.091402}{{\em Phys. Rev.
  Lett.} {\bfseries 131} no.~9, (2023) 091402},
  \href{http://arxiv.org/abs/2303.07358}{{\ttfamily arXiv:2303.07358
  [hep-th]}}.

\bibitem{Brito:2015oca}
R.~Brito, V.~Cardoso, and P.~Pani, ``{Superradiance}: {New Frontiers in Black
  Hole Physics},'' \href{http://dx.doi.org/10.1007/978-3-319-19000-6}{{\em
  Lect. Notes Phys.} {\bfseries 906} (2015) pp.1--237},
  \href{http://arxiv.org/abs/1501.06570}{{\ttfamily arXiv:1501.06570 [gr-qc]}}.

\bibitem{Guica:2008mu}
M.~Guica, T.~Hartman, W.~Song, and A.~Strominger, ``{The Kerr/CFT
  Correspondence},'' \href{http://dx.doi.org/10.1103/PhysRevD.80.124008}{{\em
  Phys. Rev. D} {\bfseries 80} (2009) 124008},
  \href{http://arxiv.org/abs/0809.4266}{{\ttfamily arXiv:0809.4266 [hep-th]}}.

\bibitem{Ceplak:2023afb}
N.~\v{C}eplak, R.~Emparan, A.~Puhm, and M.~Toma\v{s}evi\'c, ``{The
  correspondence between rotating black holes and fundamental strings},''
  \href{http://arxiv.org/abs/2307.03573}{{\ttfamily arXiv:2307.03573
  [hep-th]}}.

\bibitem{Graham:2007va}
N.~Graham and K.~D. Olum, ``{Achronal averaged null energy condition},''
  \href{http://dx.doi.org/10.1103/PhysRevD.76.064001}{{\em Phys. Rev. D}
  {\bfseries 76} (2007) 064001},
  \href{http://arxiv.org/abs/0705.3193}{{\ttfamily arXiv:0705.3193 [gr-qc]}}.

\bibitem{Emparan:1999pm}
R.~Emparan, C.~V. Johnson, and R.~C. Myers, ``{Surface terms as counterterms in
  the AdS / CFT correspondence},''
  \href{http://dx.doi.org/10.1103/PhysRevD.60.104001}{{\em Phys. Rev. D}
  {\bfseries 60} (1999) 104001},
  \href{http://arxiv.org/abs/hep-th/9903238}{{\ttfamily arXiv:hep-th/9903238}}.

\bibitem{carroll_2019}
S.~M. Carroll, \href{http://dx.doi.org/10.1017/9781108770385}{{\em Spacetime
  and Geometry: An Introduction to General Relativity}}.
\newblock Cambridge University Press, 2019.

\end{thebibliography}\endgroup
\bibliographystyle{utcaps}

\end{document}